\let\orilabel\label % Save the original kernel definition
\documentclass[reprint,amsmath,amssymb,aps,prapplied,showkeys,superscriptaddress,frontmatterverbose]{revtex4-2}
\let\label\orilabel % Force restore it over RevTeX's version

\usepackage{graphicx} % Include figure files
\usepackage{dcolumn} % Align table columns on decimal point
\usepackage{bm} % bold math
\usepackage[mathlines]{lineno}% Enable numbering of text and display math
\usepackage{multirow}
\usepackage{array}
\usepackage{makecell}
\usepackage{enumitem}
\usepackage{booktabs}
\usepackage{ragged2e}
\usepackage[dvipsnames]{xcolor}

\usepackage{amsfonts}
\usepackage{bbm}   % needed for LOWERCASE blackboard-bold (tortuosity t, 
\newcommand{\ft}[1]{\overline{#1}}          % Fourier-transformed quantity: \ft{c} -> \bar c
\newcommand{\tc}{\overline{c}}               % shorthand for the one you asked for
\newcommand{\nhat}{\widehat{\mathbf{n}}}     % outward unit normal (fluid->solid)
\newcommand{\tf}[1]{\widetilde{#1}}          % generic tilde (frequency-domain quantity)
\newcommand{\avg}[1]{\langle #1 \rangle}      % intrinsic average <.>
\newcommand{\avgt}[1]{\langle \widetilde{#1} \rangle}  % <tilde .> combo (very common)
\newcommand{\Shnd}{\avgt{Sh}}                          % nondimensional:  <Sh~>  (NOT <h~>!)

\newcommand{\xl}{\mathbf{x}}     % local (pore-scale) position vector
\newcommand{\uu}{\mathbf{u}}     % velocity field
\newcommand{\jj}{\mathbf{j}}     % species flux vector
\newcommand{\bcv}{\mathbf{b}}    % closure variable (vector, for grad-driven response)

\newcommand{\Da}{Da}                          % Damkohler number
\newcommand{\Pe}{Pe}                          % Peclet number
\newcommand{\wstar}{\omega^{*}}               % nondimensional frequency

\newcommand{\lP}{l_P}
\newcommand{\lRVE}{l_{RVE}}

\newcommand{\Vf}{V_f}
\newcommand{\Vs}{V_s}
\newcommand{\VRVE}{V_{RVE}}
\newcommand{\sA}{s_A}
\newcommand{\eps}{\varepsilon}

\newcommand{\Atot}{A_s}          % total reactive surface area (dimensional, m^2)
\newcommand{\Cnd}{\avgt{C}}              % nondimensional concentration-ratio transfer function <C~>
\newcommand{\Zeff}{\avgt{Z}}             % dimensional (area-specific) impedance <Z~>
\newcommand{\Znd}{\avgt{z}}              % nondimensional impedance <z~>
\newcommand{\iu}{i}      % imaginary unit, i^2 = -1

\usepackage[hidelinks]{hyperref} % add hypertext capabilities

\begin{document}

    \preprint{APS/123-QED}
    \title{Universal Scaling and Geometric Correction of High-Frequency Mass Transfer in Porous Electrodes}

\author{Md Abdul Hamid}
    \email{mahamid.me.buet@gmail.com}
    \affiliation{Department of Mechanical Science and Engineering, University of Illinois Urbana-Champaign, Urbana, Illinois 61801, USA}
    \affiliation{Department of Mechanical Engineering, Clemson University, Clemson, South Carolina 29634, USA}

\author{Kyle C. Smith}
    \affiliation{Department of Mechanical Science and Engineering, University of Illinois Urbana-Champaign, Urbana, Illinois 61801, USA}
    \affiliation{Department of Materials Science and Engineering, University of Illinois Urbana-Champaign, Urbana, Illinois 61801, USA}
    \affiliation{Computational Science and Engineering Program, University of Illinois Urbana-Champaign, Urbana, Illinois 61801, USA}

\date{\today}
    \begin{abstract}
Classical film theory predicts a finite diffusion-limited current at porous electrodes. Transient operation can nevertheless produce peak interfacial fluxes above this steady limit, although the time-averaged flux remains bounded. Here, we use a frequency-domain volume-averaging framework to study this transient response in porous media. In the perfectly reactive limit, the spectral Sherwood number approaches
$\Shnd\sim\sqrt{\iu\wstar}$ at high frequency, with magnitude proportional to $\omega^{1/2}$, phase $\pi/4$, and a universal leading prefactor. A boundary-layer analysis shows that this response follows from local transient diffusion and is independent of the global pore geometry. Numerical results for parallel plates, circular tubes, and twelve porous microstructures confirm the scaling across variations in particle shape, symmetry, porosity, array arrangement, flow orientation, and flow strength. We also derive the first geometry-dependent correction, which depends on interfacial area, surface curvature, and sharp edges, and verify it without fitted parameters. Comparisons with published impedance measurements for a carbon-felt fiber and a vanadium flow-battery electrode are consistent with the predicted transport response. Finally, we derive bounds on pulse duration and identify operating windows for representative redox couples. These results establish a temporal route to enhanced peak transport in porous electrodes without requiring surface charge, confinement, or electrokinetic instability.
\end{abstract}
    \maketitle
    
    \section{Introduction}

Transient operation is central to many electrochemical technologies. Pulsed and pulse-reverse currents are used in electroplating, including copper deposition for electronics manufacturing, to control deposit morphology and grain structure~\cite{Ibl1980,Despic1971,Mariani2022}. Similar protocols are used in metal recovery and refining~\cite{ZLiu2023,Tian2025,Liu2019}. Transient operation is also intrinsic to energy storage because renewable generation is intermittent and demand varies with time~\cite{Rodriguez2026,Li2019}. Redox-flow batteries (RFBs) are well suited to long-duration, large-scale storage because they decouple energy capacity from power~\cite{Soloveichik2015,Weber2011}. Vanadium RFBs are the most commercially developed, while aqueous organic and nonaqueous chemistries offer tunable redox potentials~\cite{Lourenssen2019,Huskinson2014,Gong2015,Wei2014,Zhao2021,Stumme2023}.

Despite the importance of transient operation, interfacial mass transfer in porous media is commonly described using steady film theory. Under steady conditions, the diffusion boundary layer reaches a fixed thickness. This thickness determines a finite Sherwood number, which measures interfacial mass transfer relative to molecular diffusion, and imposes a finite diffusion-limited current density. Redox-flow batteries and other electrochemical technologies that use porous electrodes are therefore usually operated below this limit to reduce concentration polarization~\cite{Weber2011}.

Currents above the diffusion-limited value, known as over-limiting currents (OLCs), have nevertheless been observed at ion-selective membranes and in microchannels, and are used in electrodialysis~\cite{Dydek2011,Nam2015,Nikonenko2014}. Existing explanations generally modify steady-state transport. Proposed mechanisms include surface conduction, electro-osmotic flow, electro-osmotic instability, geometric field focusing, and water-splitting-driven charge generation~\cite{Rubinstein1979,Rubinstein2000,Zaltzman2007,Dukhin1991,Mishchuk1995,Dydek2011,Deng2013,Han2014,Nam2015,Urtenov2013,Yaroshchuk2012,Kharkats1985,Nikonenko2014,Yossifon2010}. These mechanisms typically require surface charge, geometric confinement, additional charge carriers, or field-driven flow~\cite{Dydek2011}.

Transient operation provides a different route to enhanced interfacial transport. During a short current pulse, the depletion layer has not yet reached its steady thickness. The instantaneous current can therefore exceed the steady diffusion-limited value, even though the time-averaged current remains bounded by that value~\cite{Cheh1971,Ibl1980,Hamid2023}. The associated mass-transfer response is described by the Warburg impedance~\cite{Warburg1899,BardFaulkner2001}, which is routinely used in electroanalytical chemistry. In AC admittance voltammetry, the Warburg contribution is separated from the charge-transfer resistance through its characteristic $45^\circ$ phase. The charge-transfer resistance can then be used to determine the reaction rate constant. For very fast redox couples, this separation requires frequencies near 100~kHz, where the two contributions become comparable~\cite{Fawcett1994,Baranski1991,Muzikar2004}.

These observations motivate a pore-scale description of transient interfacial transport. In lithium-ion electrodes, salt depletion in the electrolyte can limit transport during fast charging~\cite{Liu2019b,Meyer2025}. Operando studies have also revealed heterogeneous reaction and transport fields inside working porous flow-battery electrodes~\cite{Wong2021}. Such measurements can improve electrode-scale modeling and guide microstructure design~\cite{Graf2024}. Incorporating pore-scale information into electrode-scale models requires an upscaling framework. Phenomenological models introduce constitutive relations for volume-averaged fluxes and effective transport coefficients and calibrate them against experiments or pore-scale simulations. Bottom-up models instead derive these relations from pore-scale conservation laws and interfacial boundary conditions by solving closure problems. The closure solutions determine the effective coefficients and their dependence on frequency, flow, and interfacial reactivity. Our framework follows this bottom-up approach~\cite{PRE_companion,Hamid2023}.

We previously showed that short-duration current pulses can produce redox-flow-battery current densities above the classical film-theory prediction~\cite{Hamid2023}. That formulation, however, assumed a spatially uniform macroscopic concentration. A subsequent frequency-domain volume-averaging framework removed this assumption and introduced the spectral Sherwood number, the frequency-dependent form of the Sherwood number~\cite{PRE_companion}. In the limit of a perfectly reactive interface, earlier studies found that this quantity grows with the square root of frequency and has a phase of $45^\circ$ for cylindrical pores~\cite{Hamid2023} and for parallel plates, circular tubes, and a periodic array of square rods~\cite{PRE_companion}.

It remains unclear whether this scaling holds for realistic porous microstructures and flow conditions, how rapidly different microstructures approach the high-frequency limit, and which chemistries and electrodes can access this regime. These issues are important because measured standard rate constants span more than four orders of magnitude, from vanadium couples on carbon~\cite{Yamamura2005} to quinones~\cite{Huskinson2014}, TEMPO ((2,2,6,6-tetramethylpiperidin-1-yl)oxyl) catholytes~\cite{Suga2004,Stumme2023}, and ferrocene and hexaammineruthenium at well-characterized metal electrodes~\cite{Sun2006,Muzikar2004}.

Here, we study twelve porous microstructures that differ in particle shape, symmetry, porosity, array arrangement, flow orientation, and flow strength. We use the P\'eclet number to characterize the relative importance of advection and diffusion. By resolving the frequency-dependent response, we determine the crossover from pseudo-steady to high-frequency transport and quantify how rapidly different microstructures approach the asymptotic limit. We show that, for perfectly reactive interfaces, the high-frequency spectral Sherwood number follows the same square-root frequency scaling, with a $45^\circ$ phase, for all geometries and flow conditions considered. A boundary-layer analysis shows that this leading response is determined by local transient diffusion near the interface rather than by the global geometry of the porous electrode.

We also derive, through an asymptotic boundary-layer analysis, the first geometry-dependent correction to the high-frequency response for microstructures with smooth surfaces and sharp edges. The correction depends on interfacial area, porosity, surface curvature, and edge angles. Without fitted parameters, it predicts the simulated correction for eleven two-dimensional microstructures and agrees with an exact solution for spheres. Related short-time expansions are known for restricted diffusion in porous media~\cite{Mitra1992,Mitra1993} and heat flow from polygons~\cite{vandenBerg1990}. We obtain the corresponding result for the closure problem and the spectral Sherwood number, using the polygonal heat-flow result to determine the contribution of sharp edges.

We compare the predictions with published impedance measurements for a carbon fiber and a flow-battery electrode~\cite{LandonLane2019,Vivona2020}. Finally, we derive bounds on the time-averaged flux, peak flux, and pulse duration and evaluate the resulting operating window for specific redox couples. These results identify when high-frequency interfacial transfer can be accessed in aqueous and nonaqueous flow batteries, impedance diagnostics, and pulsed electrodeposition. The work establishes a temporal route to peak interfacial fluxes above the steady limit without requiring surface charge, confinement, additional charge carriers, or electrokinetic instability.
    \section{Theory}

\subsection{Pore-scale model and closure}

Consider a solute with molecular diffusivity $D$ transported by a steady, incompressible flow $\uu$ through a periodic porous medium. The solute reacts at the solid--liquid interface $s$. Its Fourier-transformed concentration $\tc(\xl,\omega)=\mathcal{F}\{c(\xl,t)\}$ satisfies
\begin{equation}
    \iu \omega \tc +\nabla\cdot(\uu \tc-D\nabla \tc)=0,
    \label{eq:pore1}
\end{equation}
\begin{equation}
    \nhat\cdot D\nabla \tc\big|_s=k\,(\tc_{s,eq}-\tc_{s}),
    \label{eq:pore2}
\end{equation}
where $\iu=\sqrt{-1}$, $\nhat$ is the unit normal to the interface pointing from the fluid into the solid, $k$ is the interfacial reaction-rate constant, $\tc_s$ is the concentration at the interface, and $\tc_{s,eq}$ is the interfacial concentration in equilibrium with the electrode potential.

We perform volume averaging over a representative volume element (RVE), defined as a periodic unit cell of volume $\VRVE$. The RVE contains fluid volume $\Vf$, porosity $\eps=\Vf/\VRVE$, and solid--liquid interfacial area $\Atot$. The interfacial area per unit RVE volume is $\sA=\Atot/\VRVE$, and $\avg{\cdot}$ denotes the intrinsic fluid-phase average. The concentration fluctuation is represented by two closure variables~\cite{PRE_companion},
\begin{equation}
    \tc-\avg{\tc}=a(\xl,\omega)\,(\tc_{s,eq}-\avg{\tc})+\bcv(\xl,\omega)\cdot\nabla\avg{\tc}.
    \label{eq:closure}
\end{equation}
This representation follows from the linearity of the pore-scale transport problem. It separates the response to two independent macroscopic driving effects: the concentration difference between the equilibrium interface and the fluid-phase average, and the macroscopic concentration gradient. The scalar $a$ describes the first response, while the vector $\bcv$ describes the second. Both variables are periodic over the RVE, but only $a$ enters the interfacial transfer coefficient considered here.

The closure problem for $a$ is
\begin{equation}
    \begin{aligned}
        &\iu\omega a+\nabla\cdot(\uu a-D\nabla a)=\alpha,\\
        &\left.\left(D\nhat\cdot\nabla a+ka\right)\right|_s=k.
    \end{aligned}
    \label{eq:a_closure}
\end{equation}
The first equation is obtained by substituting the closure representation into the Fourier-transformed pore-scale transport equation and collecting the terms proportional to the interfacial concentration difference. The Robin condition follows from the pore-scale interfacial condition in Eq.~\eqref{eq:pore2}. The spatially uniform source $\alpha$ enforces the zero-mean constraint $\avg{a}=0$~\cite{PRE_companion}. It is therefore a mathematical Lagrange multiplier rather than an additional physical source of solute.

The Damk\"ohler number $\Da=k\lP/D$ compares the interfacial reaction rate with diffusion over the characteristic pore length $\lP$. The limits $\Da=0$ and $\Da\to\infty$ correspond to non-reactive and perfectly reactive interfaces, respectively. The dimensionless frequency is $\wstar=\omega\lP^2/D$, the ratio of the pore-diffusion time $\lP^2/D$ to the forcing time scale $1/\omega$. The spectral Sherwood number is
\begin{equation}
    \Shnd=\frac{\eps\lP}{\sA \Vf}\int_s\nhat\cdot\nabla a\,dA,
    \label{eq:Sh}
\end{equation}
and relates the interface-averaged flux, defined as positive from the interface into the fluid, to the concentration difference through $\avg{\ft{j}_s}=(D/\lP)\,\Shnd\,(\tc_{s,eq}-\avg{\tc})$. At a fixed concentration difference, the interfacial flux and electrochemical current are therefore proportional to $\Shnd$. In the zero-frequency limit, $\Shnd$ reduces to the real pseudo-steady Sherwood number $\avg{Sh}$, which is written without the tilde. At finite frequency, $\Shnd$ is complex. Its magnitude measures the strength of interfacial transfer, and its phase $\theta_{\Shnd}$ measures the phase of the flux relative to the concentration difference. A positive phase means that the flux leads the concentration difference.

\subsection{Canonical geometries and the electrochemical impedance}

Electrochemical impedance is the ratio of the voltage response to an imposed oscillatory current. When species transport limits the response, the diffusion impedance relates the interfacial concentration perturbation to the flux required to sustain it and converts this transport response into a voltage through the dependence of electrode potential on surface concentration~\cite{Jacobsen1995}.

In the present framework, $\Shnd$ determines the interfacial flux, while solute accumulation in the pore volume produces a storage contribution. For a redox couple with negligible migration, Appendix~\ref{ap:A} gives
\begin{equation}
    \Znd=\frac{1}{\Shnd}+\frac{\sA\lP}{\eps\,\iu\wstar},
    \label{eq:ZSh}
\end{equation}
where $\Znd$ is the nondimensional diffusion impedance of one redox species, normalized by $|\nu|\lP RT/(F^2Dc^0_{s,eq})$. Here, $\nu$ is the signed stoichiometric coefficient, $R$ is the gas constant, $T$ is the temperature, $F$ is Faraday's constant, and $c^0_{s,eq}$ is the initial equilibrium interfacial concentration. The first term is the interfacial mass-transfer resistance, and the second is the storage contribution. The latter assumes a locally uniform macroscopic concentration, so that advection does not enter the RVE balance. We represent $\Znd$ using a Nyquist plot of $-\mathrm{Im}\,\Znd$ against $\mathrm{Re}\,\Znd$.

For a redox couple, the total impedance is the sum of the oxidized- and reduced-species contributions; the normalization in Eq.~\eqref{eq:ZSh} applies to one species. Impedance spectra can also distinguish local interfacial behavior from transport across the full electrode. For a rotating-disk electrode with a non-uniformly reactive surface, the low-frequency impedance reflects the entire electrode, whereas the high-frequency impedance samples an isolated active site~\cite{Caprani1987}. The frequency separating these regimes provides information about the active-region size. The same separation between local and global transport scales occurs in the present system.

\begin{figure*}[ht!]
    \includegraphics[width=0.98\linewidth]{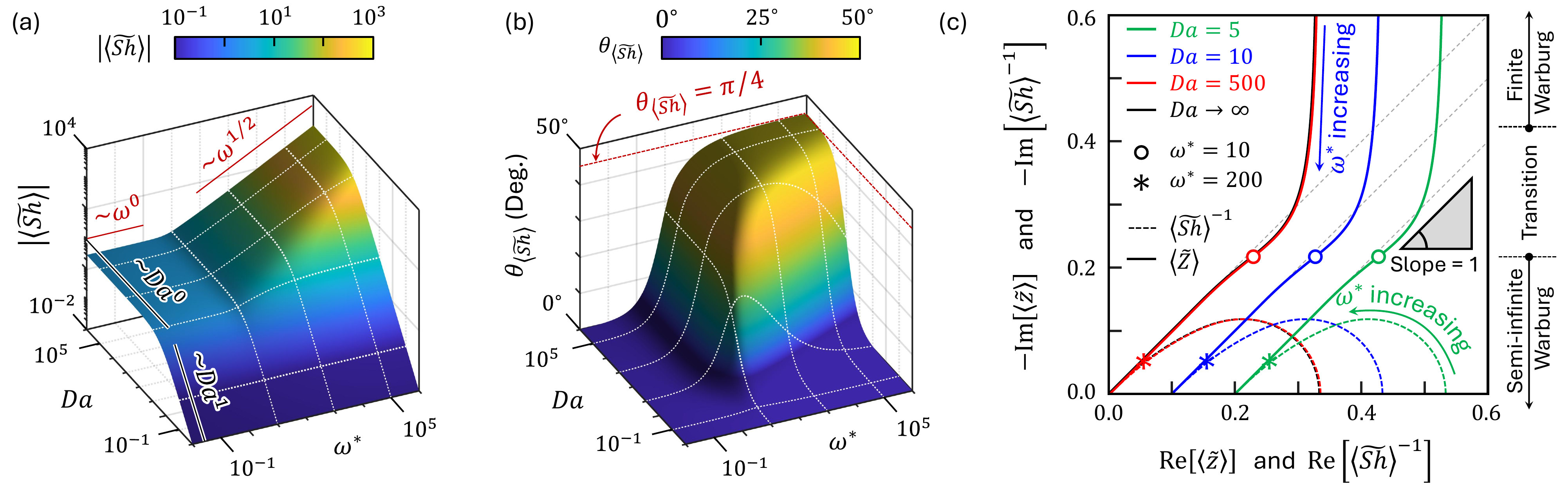}
    \caption{(a,b)~Magnitude and phase of the spectral Sherwood number $\Shnd$ as functions of the Damk\"ohler number $\Da$ and dimensionless frequency $\wstar$ for Poiseuille flow between parallel plates. At low frequency, $|\Shnd|$ is proportional to $\Da$ in the reaction-limited regime and independent of $\Da$ in the diffusion-limited regime. At high frequency and large $\Da$, $|\Shnd|\sim\omega^{1/2}$ and the phase approaches $\pi/4$. (c)~Nyquist plots of the nondimensional impedance $\Znd$ (solid) and $1/\Shnd$ (dashed) for $\Da=5$, $10$, $500$, and $\Da\to\infty$. Circles and asterisks mark $\wstar=10$ and $200$. The labels identify the semi-infinite-Warburg, transition, and finite-Warburg regimes.}
    \label{fig:1}
\end{figure*}

For fully developed Poiseuille flow between parallel plates of half-spacing $h$ and through circular tubes of radius $r_t$, the closure problem is one-dimensional and has analytical solutions. Setting $\lP=h$ for plates and $\lP=r_t$ for tubes, Eq.~\eqref{eq:ZSh} gives
\begin{equation}
    \Znd=\frac{1}{\Da}+\dfrac{\cosh(\sqrt{\iu \wstar})}{(\sqrt{\iu \wstar})(\sinh\sqrt{\iu \wstar})}, \qquad \text{(plates)}
    \label{eq:z_Plates}
\end{equation}
\begin{equation}
    \Znd=\frac{1}{\Da}+\dfrac{I_0(\sqrt{\iu \wstar})}{(\sqrt{\iu \wstar})\,I_1(\sqrt{\iu \wstar})}, \qquad \text{(tube)}
    \label{eq:z_Tube}
\end{equation}
where $I_0$ and $I_1$ are modified Bessel functions of first kind. The second terms are the finite-length diffusion impedances for planar and cylindrical geometries with an impermeable far boundary~\cite{Jacobsen1995}; the first term is the finite-kinetics resistance.

Figure~\ref{fig:1}(a,b) shows $|\Shnd|$ and $\theta_{\Shnd}$ for parallel plates. At low frequency, the concentration field adjusts across the entire gap and the response is pseudo-steady. The magnitude is proportional to $\Da$ in the reaction-limited regime and independent of $\Da$ in the diffusion-limited regime. As frequency increases, the disturbance becomes confined near the interface. For large $\Da$, the transition begins at $\wstar\sim10$, after which $|\Shnd|$ grows as $\omega^{1/2}$ and the phase approaches $\pi/4$.

Figure~\ref{fig:1}(c) shows the same transition in the Nyquist plane. For $\Da\to\infty$, the $\Znd$ curve leaves the origin along a $45^\circ$ line, representing the semi-infinite Warburg response. For finite $\Da$, the high-frequency response is shifted along the real axis by $1/\Da$. At lower frequencies, the curve bends through a transition region and approaches a nearly vertical finite-Warburg response as the disturbance samples the entire pore. The dashed $1/\Shnd$ curves differ from $\Znd$ only by the storage term in Eq.~\eqref{eq:ZSh}; they therefore coincide at high frequency and separate at low frequency. At low frequency, $1/\Shnd$ closes on the real axis at $1/\avg{Sh}=1/3+1/\Da$ for parallel plates. At high frequency, both quantities approach $1/\Da$ for finite $\Da$ and the origin for $\Da\to\infty$.

At high frequency, the storage term decays as $\omega^{-1}$, faster than $1/\Shnd$. Therefore, in the diffusion-limited limit, $\Znd\approx1/\Shnd$ and
\begin{equation}
    |\Shnd|\sim\omega^{1/2},\qquad\theta_{\Shnd}\to\pi/4\qquad(\Da\to\infty).
    \label{eq:scaling}
\end{equation}
This unbounded growth occurs only in the diffusion-limited limit. At finite $\Da$, the kinetic resistance remains finite and frequency-independent. Once $\wstar\gtrsim\Da^2$, it dominates the decreasing diffusive resistance, causing $\Shnd$ to approach the finite limit $\Da$ and $\theta_{\Shnd}$ to return toward zero.
    \section{High-frequency asymptotics}

\subsection{Leading-order scaling}

The parallel-plate results suggest a universal $\Shnd$--$\wstar$ scaling at high frequency. At angular frequency $\omega$, the concentration disturbance penetrates into the fluid over the diffusion length $\delta=\sqrt{D/\omega}$. When $\delta\ll\lP$, every smooth interface appears locally planar and adjacent to a semi-infinite fluid domain. We derive the high-frequency exponent and leading prefactor directly from the closure problem.

For $\Da\to\infty$, the interface is perfectly reactive and $a=1$ there. Because the flow is incompressible, Eq.~\eqref{eq:a_closure} becomes
\begin{equation}
    \iu\omega a+\uu\cdot\nabla a=D\nabla^2a+\alpha,\qquad a\big|_s=1.
    \label{eq:B1}
\end{equation}
Away from the interface, $a$ approaches the uniform value $a_\infty=\alpha/(\iu\omega)$. The zero-mean constraint gives $a_\infty=O(\delta/\lP)$, so it is negligible at leading order. Let $n$ denote the distance from the interface into the fluid and define $\eta=n/\delta$. The normal-diffusion term scales as $D/\delta^2$, while the temporal term scales as $\omega$. Thus, $\delta=\sqrt{D/\omega}$ makes these terms comparable.

The no-slip condition sets the tangential velocity to zero at the interface, while impermeability sets the normal velocity to zero. A Taylor expansion in $n$ therefore gives $u_\tau=O(n)$. Because the tangential velocity vanishes along the entire interface, its tangential derivative also vanishes there. Incompressibility then requires the first normal derivative of the normal velocity to vanish at the interface, giving $u_n=O(n^2)$. Consequently, advection is of relative order $\Pe(\delta/\lP)^3=\Pe/(\wstar)^{3/2}$, and its contribution to $\Shnd$ is of order $\Pe/\wstar$. Curvature contributes at relative order $\delta/\lP$, while tangential diffusion contributes at order $(\delta/\lP)^2$. The leading-order boundary-layer problem is therefore
\begin{equation}
    \frac{\partial^2a}{\partial\eta^2}=\iu a,\qquad a(0)=1,\qquad a\to0\ \ (\eta\to\infty).
    \label{eq:B2}
\end{equation}
The solution is $a=\exp(-\sqrt{\iu}\,\eta)$, and the leading wall-normal gradient is $\nhat\cdot\nabla a=\sqrt{\iu\omega/D}$. Substitution into Eq.~\eqref{eq:Sh} gives
\begin{equation}
    \Shnd=\lP\sqrt{\frac{\iu\omega}{D}}+O(1)=\sqrt{\iu\wstar}+O(1).
    \label{eq:B3}
\end{equation}

Equation~\eqref{eq:B3} shows that the exponent and leading prefactor are universal. The prefactor is unity because $\Shnd$ and $\wstar$ use the same characteristic length $\lP$. In dimensional form, the mass-transfer coefficient is $\sqrt{\iu\omega D}$ and contains no pore length. Since the storage term in Eq.~\eqref{eq:ZSh} scales as $(\wstar)^{-1}$, $\Znd\approx{\Shnd}^{-1}\sim(\wstar)^{-1/2}$, giving the semi-infinite Warburg response~\cite{Warburg1899,BardFaulkner2001}.

\subsection{First-order correction for an arbitrary solid}
\label{sec:correction}

Equation~\eqref{eq:B3} leaves the $O(1)$ term undetermined, although this term controls how rapidly a microstructure approaches the high-frequency limit. We write
$a=a_\infty+(1-a_\infty)\Phi$, where $\Phi$ decays away from the interface, to separate the storage and geometric contributions (Appendix~\ref{ap:B}). The zero-mean constraint gives $a_\infty\approx-\sA\lP/(\eps\sqrt{\iu\wstar})$, which increases the concentration difference across the boundary layer. The remaining correction depends on whether a surface feature appears smooth or sharp over the diffusion length $\delta=\sqrt{D/\omega}$.

We call a surface feature smooth when its radius of curvature $R_c$ satisfies $R_c\gg\delta$ and sharp when $R_c\ll\delta$. Because $\delta=\lP/\sqrt{\wstar}$, this classification depends on frequency. A true corner, for which $R_c=0$, is sharp at every frequency. A rounded corner appears sharp at low frequency and smooth at sufficiently high frequency. Features with $R_c\sim\delta$ are transitional and are not described by either limiting case.

Near a smooth surface,
$\nabla^2\Phi\approx\partial_n^2\Phi+2\rho\partial_n\Phi$, where $\rho=(\kappa_1+\kappa_2)/2$ is the local mean curvature and $\kappa_1$ and $\kappa_2$ are the principal curvatures. We take $\rho>0$ where the solid is convex, that is, where it bulges into the fluid. Convex surfaces therefore increase the wall gradient by $\rho$, whereas concave surfaces, such as the inside of a tube, decrease it. Near a sharp edge, the boundary layer follows the solution for a fluid wedge with angle $\gamma$, measured through the fluid. This angle is related to the interior solid angle $\beta$ by $\gamma=2\pi-\beta$. A flat surface has $\gamma=\pi$, a convex solid corner has $\gamma>\pi$, and a reentrant solid corner has $\gamma<\pi$. Combining the storage, curvature, and edge contributions gives
\begin{equation}
    \begin{aligned}
        &\Shnd=\sqrt{\iu\wstar}+C+O\!\left((\iu\wstar)^{-1/2}\right)+O\!\left(\frac{\Pe}{\wstar}\right),\\
        &C=\frac{\sA\lP}{\eps}+\lP\avg{\rho}_s-\frac{\lP}{\Atot}\sum_e \Lambda(\gamma_e)L_e,
    \end{aligned}
    \label{eq:C_general}
\end{equation}
where $\avg{\rho}_s=\Atot^{-1}\int\rho\,dA$ is the area average over the smooth part of the surface. The sum runs over the sharp edges, $L_e$ is the length of edge $e$, and $\gamma_e$ is its fluid-side angle. The three terms in $C$ represent storage in the pore volume, curvature of the smooth surface, and sharp edges, respectively. A surface may contain both smooth and sharp portions, in which case both geometric terms are included. The edge coefficient is~\cite{vandenBerg1990}
\begin{equation}
    \Lambda(\gamma)=\int_0^\infty\frac{4\sinh[(\pi-\gamma)\zeta]}{\sinh(\pi \zeta)\cosh(\gamma \zeta)}\,d\zeta,\qquad 0<\gamma<2\pi,
    \label{eq:Lambda_gamma}
\end{equation}
with $\Lambda(\pi)=0$, $\Lambda(\pi/2)=4/\pi$, and $\Lambda(3\pi/2)\approx-0.602$. Because $\Lambda<0$ for $\gamma>\pi$ and $\Lambda>0$ for $\gamma<\pi$, convex solid corners increase $C$, whereas reentrant corners decrease it. Vertices of three-dimensional solids contribute at higher order and are not included. The result applies in both two and three dimensions and contains no fitted parameters. It requires
\begin{equation}
    \wstar\gg\max\left[\avg{Sh}^2,\;(\lP/g_{\min})^2,\;\Pe\right],
    \label{eq:C_conditions}
\end{equation}
where $g_{\min}$ is the narrowest local pore dimension. The last condition is stricter than the condition $\wstar\gg\Pe^{2/3}$ required for the leading term because the advective contribution to $\Shnd$, of order $\Pe/\wstar$, must be small compared with the $O(1)$ correction rather than only with the leading term $\sqrt{\wstar}$.

The constant $C$ is independent of the P\'eclet number and flow direction. Flow enters $\Shnd$ through the term of order $\Pe/\wstar$, which vanishes at high frequency because the no-slip and impermeability conditions make the fluid nearly stagnant within the boundary layer. During one forcing period, the flow therefore displaces fluid within the boundary layer by only a small distance compared with $\lP$. It cannot transport solute into or out of the layer before the forcing reverses. The three terms in $C$ are likewise independent of the flow. The storage term follows from the uniform far-field value $a_\infty$, which is fixed by the constraint $\avg{a}=0$, while the curvature and edge terms depend only on the local surface geometry. Flow nevertheless affects the approach to the high-frequency limit by setting the pseudo-steady Sherwood number $\avg{Sh}$ and, consequently, the crossover frequency.

In two dimensions, the solid is a prismatic inclusion of infinite length. All quantities are taken per unit depth, so $\Atot$ is the perimeter of the inclusion, each corner is an edge with $L_e=1$, and $\rho=\kappa/2$, where $\kappa$ is the in-plane boundary curvature. For $\lP=d_P=4V_s/\Atot$, the storage term is $\sA\lP/\eps=4(1-\eps)/\eps$ for any inclusion shape, whereas the geometric term depends on whether the boundary is smooth or sharp. For a single closed inclusion per unit cell with an everywhere-smooth boundary, the planar curvature satisfies $\int\kappa\,ds=2\pi$, and therefore $\int\rho\,ds=\pi$. Equation~\eqref{eq:C_general} then reduces to
\begin{equation}
    C=\frac{4(1-\eps)}{\eps}+\frac{\pi d_P}{\Atot}=\frac{4(1-\eps)}{\eps}+\frac{4\pi V_s}{\Atot^2}.
    \label{eq:C_2D}
\end{equation}
The second term is a shape factor equal to 1 for a circle and less than 1 for any other smooth shape. Equation~\eqref{eq:C_2D} applies only to smooth closed inclusions. It does not apply to polygons, for which the geometric contribution is the edge sum in Eq.~\eqref{eq:C_general} with $\avg{\rho}_s=0$, or to inclusions combining arcs and corners, for which both geometric terms must be evaluated. For $M$ separate smooth inclusions per unit cell, $\pi$ in Eq.~\eqref{eq:C_2D} is replaced by $M\pi$, with $\Atot$ denoting the total perimeter. Solids that connect across the unit cell, such as parallel plates, are not closed inclusions, so their curvature must be integrated directly.

In three dimensions, $\int\rho\,dA$ is not topological and depends on shape. For spheres of diameter $d$ with $\lP=d$, $\lP\rho=2$, and Eq.~\eqref{eq:C_general} gives $C=6(1-\eps)/\eps+2$. For straight fibers with $\lP=d$, $\lP\rho=1$, giving $C=4(1-\eps)/\eps+1$. For a smooth convex solid, $\int\rho\,dA=2\pi l_w$, where $l_w$ is the mean width. In general, this integral can be evaluated from a surface mesh or tomographic reconstruction.

Equation~\eqref{eq:C_general} has two direct consequences. First, the approach to the Warburg limit follows
\begin{equation}
    \frac{|\Shnd|}{\sqrt{\wstar}}\approx1+\frac{C}{\sqrt{2\wstar}},\qquad\theta_{\Shnd}\approx\frac{\pi}{4}-\frac{C}{\sqrt{2\wstar}}.
    \label{eq:approach}
\end{equation}
Second, the storage contribution in $C$ cancels the explicit storage term in Eq.~\eqref{eq:ZSh}. Since
$C=\sA\lP/\eps+C_{\mathrm{geom}}$, where $C_{\mathrm{geom}}$ contains the curvature and edge contributions, combining Eqs.~\eqref{eq:ZSh} and \eqref{eq:C_general} gives
\begin{equation}
    \Znd\approx\frac{1}{\sqrt{\iu\wstar}}+\frac{\sA\lP/\eps-C}{\iu\wstar}.
    \label{eq:z_correction}
\end{equation}
Thus, the first correction to $\Znd$ contains only the curvature and edge contributions. The first correction to $\Shnd$, however, also contains the porosity-dependent storage contribution. Equation~\eqref{eq:z_correction} recovers the $\Da\to\infty$ expansions of Eqs.~\eqref{eq:z_Plates} and \eqref{eq:z_Tube}.

For a finite slip velocity, the tangential velocity does not vanish at the interface, and advection is of relative order $\Pe(\delta/\lP)^2$ rather than $\Pe(\delta/\lP)^3$. Its contribution to $\Shnd$ therefore scales as $\Pe/\sqrt{\wstar}$ rather than $\Pe/\wstar$. Slip leaves the exponent, leading prefactor, and constant $C$ unchanged, but it requires $\wstar\gg\Pe^2$, rather than $\wstar\gg\Pe$, for Eq.~\eqref{eq:C_general} to apply.
    \section{Porous microstructures and numerical method}

\begin{figure}[t]
    \includegraphics[width=0.97\linewidth]{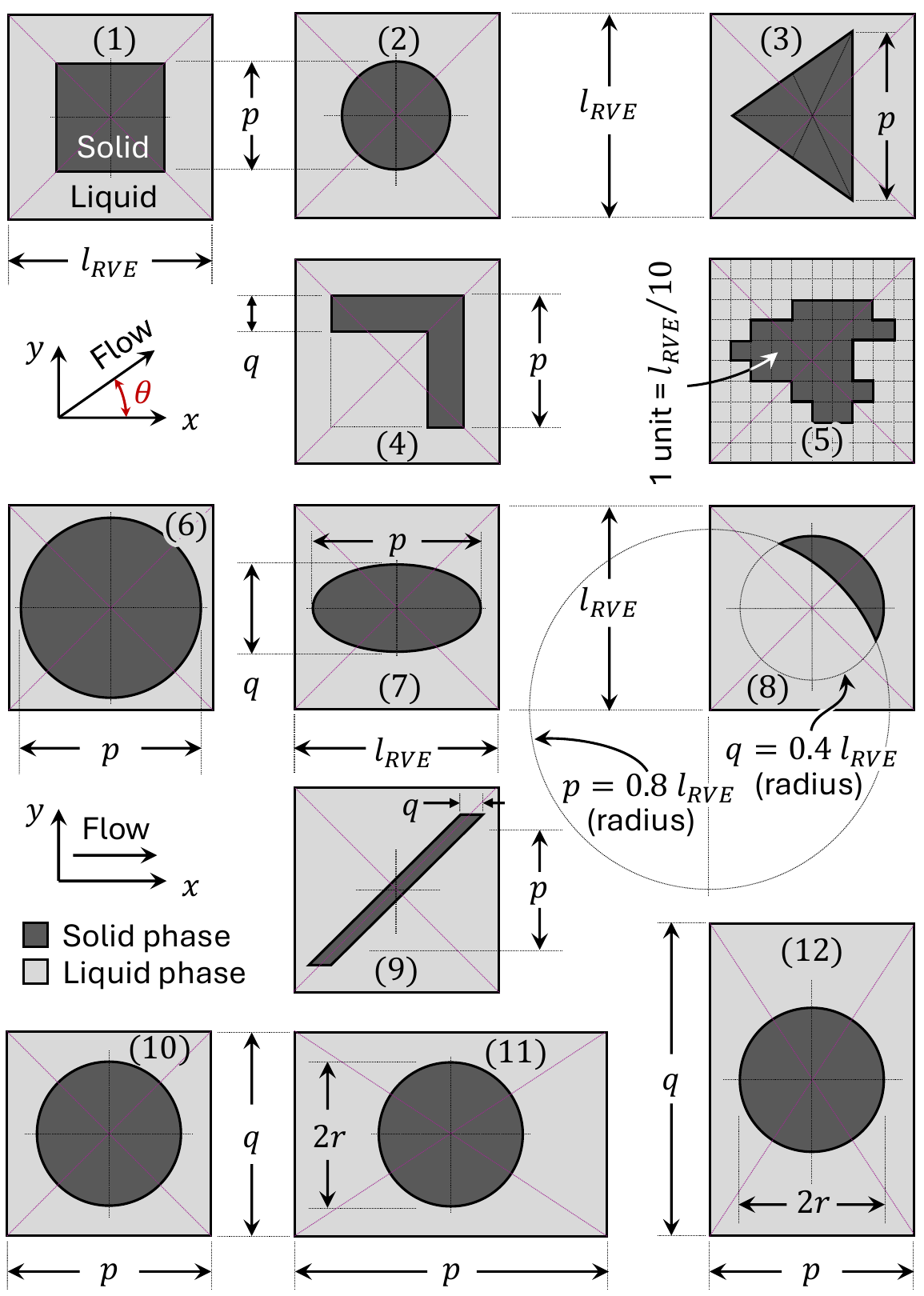}
    \caption{Periodic unit cells of the twelve porous microstructures. Dark and light regions denote the solid and liquid phases, respectively. Geometries (1)--(5) differ in particle shape and are studied for flow angles $\theta=0^\circ$, $30^\circ$, and $45^\circ$. Geometries (6)--(9) differ in porosity, while geometries (10)--(12) differ in array arrangement and RVE aspect ratio. Geometries (6)--(12) are studied only for $\theta=0^\circ$. The geometric parameters and equivalent particle diameters are listed in Table~\ref{tab:1}.}
    \label{fig:2}
\end{figure}

\begin{table}[t!]
    \centering
    \caption{Geometric details of the porous microstructures shown in Fig.~\ref{fig:2}. The definitions of $p$ and $q$ for each geometry are shown in Fig.~\ref{fig:2}. For geometries (1)--(9), the reference length is $l_{\text{ref}}=l_{\text{RVE}}$, and the RVE is a $25~\mu\mathrm{m}\times25~\mu\mathrm{m}$ square. For geometries (10)--(12), $l_{\text{ref}}=8.33~\mu\mathrm{m}$, the radius of the circular inclusion. The RVE for geometry (10) is a $25~\mu\mathrm{m}\times25~\mu\mathrm{m}$ square. The RVEs for geometries (11) and (12) are rectangles measuring $35~\mu\mathrm{m}\times25~\mu\mathrm{m}$ and $25~\mu\mathrm{m}\times35~\mu\mathrm{m}$, respectively, and differ only in their orientation relative to the flow. The characteristic length is the equivalent particle diameter, $\lP=d_P=4V_s/A_s$, where $V_s$ and $A_s$ are the solid cross-sectional area and perimeter per unit depth, respectively.}
    \label{tab:1}
    \begin{tabular*}{\columnwidth}{@{\extracolsep{\fill}}ccccc}
        \toprule
        ID & $p/l_{\text{ref}}$ & $q/l_{\text{ref}}$ & $\eps$ & $d_P$ ($\mu\mathrm{m}$) \\
        \midrule
        G-1  & 0.53 & --  & 0.720 & 13.3 \\
        G-2  & 0.6  & --  & 0.717 & 15.0 \\
        G-3  & 0.8  & --  & 0.720 & 11.5 \\
        G-4  & 0.8  & 0.2 & 0.720 & 8.8  \\
        G-5  & --   & --  & 0.720 & 9.3  \\
        \midrule
        G-6  & 0.9  & --  & 0.364 & 22.5 \\
        G-7  & 0.9  & 0.5 & 0.647 & 15.8 \\
        G-8  & 0.8  & 0.4 & 0.800 & 9.6  \\
        G-9  & 0.8  & 0.1 & 0.920 & 3.2  \\
        \midrule
        G-10 & 3.0  & 3.0 & 0.651 & 16.7 \\
        G-11 & 4.2  & 3.0 & 0.751 & 16.7 \\
        G-12 & 3.0  & 4.2 & 0.751 & 16.7 \\
        \bottomrule
    \end{tabular*}
\end{table}

We test the theory in complex porous media using twelve periodic unit cells, shown in Fig.~\ref{fig:2}. Each cell contains a two-dimensional solid inclusion representing a reactive particle. Because the inclusions extend indefinitely in the axial direction, the three-dimensional structures are represented by two-dimensional arrays of rods. Geometries (1)--(5) vary in particle shape and are studied for flow angles $\theta=0^\circ$, $30^\circ$, and $45^\circ$. Geometries (6)--(9) vary in porosity, whereas geometries (10)--(12) vary in array arrangement and RVE aspect ratio. Geometries (6)--(12) are considered only for $\theta=0^\circ$. For each geometry, the characteristic length is the equivalent particle diameter
$d_P=4V_s/A_s$, listed in Table~\ref{tab:1}, where $V_s$ and $A_s$ are the area and perimeter of the two-dimensional solid inclusion, respectively. Thus, $\lP\equiv d_P$.

The steady, incompressible, inertia-free velocity field $\uu$ is obtained by numerically solving the two-dimensional Stokes equations in stream-function form,
$\nabla^4\Psi=0$. No-slip and impermeability conditions are imposed on the rod surfaces, and periodic conditions are imposed at the cell boundaries. The imposed mean flow is represented by a constant jump in the stream function across the cell.

Flow strength is characterized by the P\'eclet number
$\Pe=|\mathbf{u}_s|d_P/D$, which is the ratio of the advective transport rate $|\mathbf{u}_s|/d_P$ to the diffusive transport rate $D/d_P^2$. Here, $\mathbf{u}_s=\eps\avg{\uu}$ is the superficial (Darcy) velocity. Values $\Pe\ll1$ and $\Pe\gg1$ correspond to diffusion- and advection-dominated transport, respectively. Because the flow is inertia-free, changing $\Pe$ rescales the velocity magnitude without altering its spatial structure. Thus, $\Pe$ isolates the effect of advection on the closure problem.

All twelve geometries are simulated at $\Pe=100$. To examine the effect of flow strength, the symmetric inclusion G-2 and asymmetric inclusion G-4 are additionally simulated at $\Pe=0.1$, $10$, $100$, and $1000$ for $\theta=0^\circ$. For each $\wstar>0$, the closure problem for $a(\xl,\omega)$ is solved using the corresponding velocity field, the Robin condition at the rod surfaces, and $k\to\infty$, corresponding to $\Da\to\infty$. Results are reported for $\wstar\leq10^4$, the range over which the solutions are grid-converged. Details of the discretization, numerical implementation, and verification are provided in our earlier work~\cite{Hamid2023} and in the companion article~\cite{PRE_companion}.
    \section{Results}

\begin{figure*}[ht!]
    \includegraphics[width=0.95\linewidth]{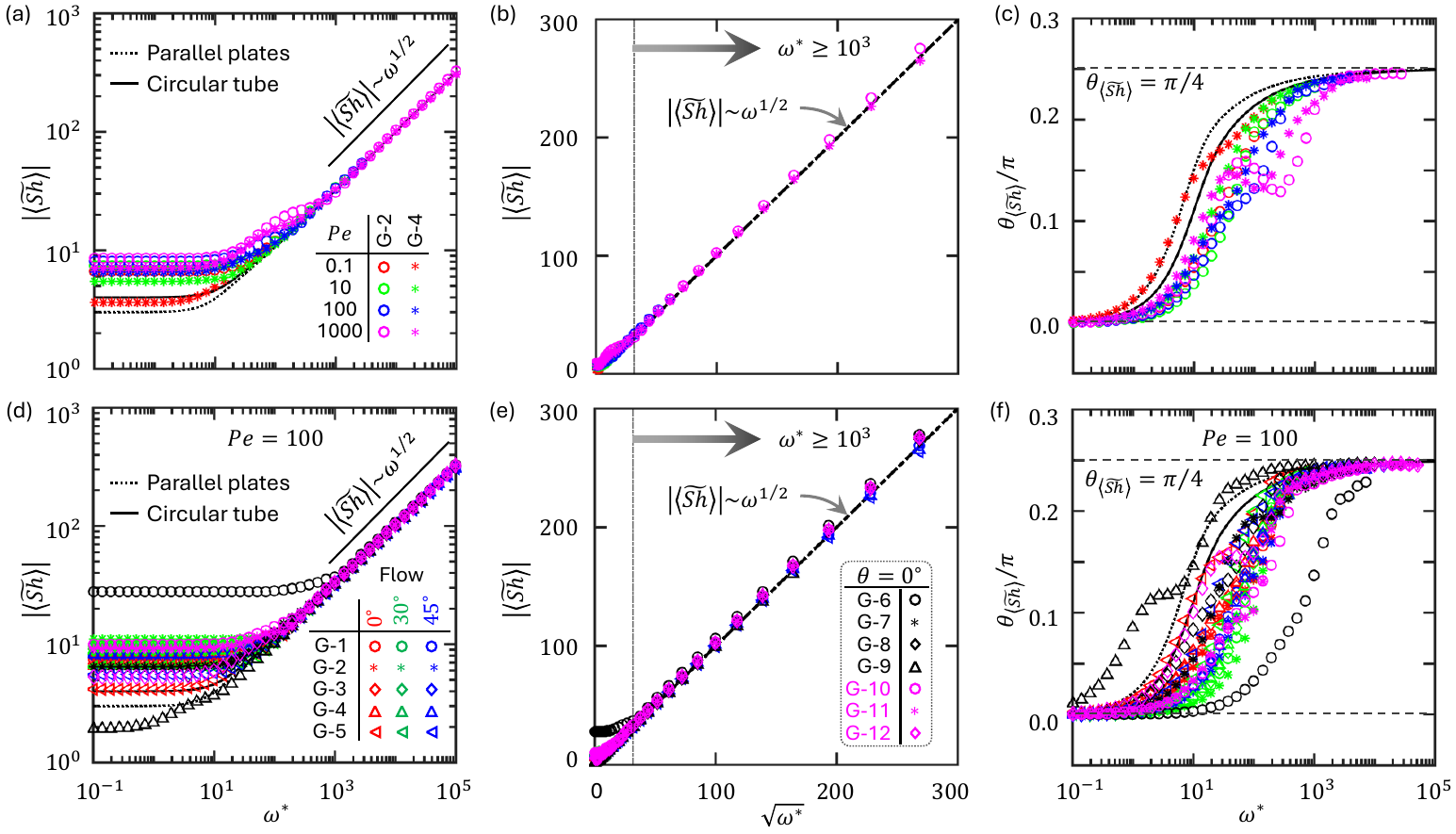}
    \caption{High-frequency scaling for $\Da\to\infty$. Left column: magnitude of $\Shnd$ versus $\wstar$ on log--log axes. Middle column: magnitude versus $\sqrt{\wstar}$ on linear axes. Right column: phase $\theta_{\Shnd}/\pi$ versus $\wstar$. Panels (a)--(c) show G-2 and G-4 at $\theta=0^\circ$ for $\Pe=0.1$, $10$, $100$, and $1000$. Panels (d)--(f) show all twelve microstructures at $\Pe=100$. The dotted and solid black curves denote the parallel-plate and circular-tube solutions, respectively. Panels in each row share a legend: the top-row legend is shown in (a), and the bottom-row legend is divided between (d) and (e). The solid slope-$1/2$ lines in (a) and (d), and the dash-dotted unit-slope lines in (b) and (e), indicate $|\Shnd|=\sqrt{\wstar}$. The vertical lines in (b) and (e) mark $\wstar=10^3$, and the horizontal dashed lines in (c) and (f) mark $\theta_{\Shnd}=0$ and $\pi/4$.}
    \label{fig:3}
\end{figure*}

\subsection{Universal high-frequency response}

All results in this section correspond to $\Da\to\infty$. Figure~\ref{fig:3} examines the high-frequency scaling across different flow strengths and porous microstructures.

At low frequency, the response is pseudo-steady and strongly dependent on the geometry. In Fig.~\ref{fig:3}(d), the plateau of $|\Shnd|$ ranges from approximately 2 for the thin inclined strip G-9 to approximately 28 for the low-porosity inclusion G-6. The corresponding plateaus for parallel plates and circular tubes, approximately 3 and 4, lie within this range.

Flow strength also affects the plateau [Fig.~\ref{fig:3}(a)]. For G-2, the plateau increases from approximately 6.6 at $\Pe=0.1$ to approximately 8.5 at $\Pe=1000$, whereas for G-4 it increases from approximately 3.6 to approximately 7.1. As $\Pe$ increases, advection redistributes solute along the interface and enhances the pseudo-steady transfer rate. This effect is stronger for the asymmetric geometry. Thus, the pseudo-steady Sherwood number, which determines the classical limiting current, depends on the microstructure, flow direction, and flow strength.

At high frequency, the effects of flow strength and flow direction become negligible, and the data approach the same curve as the canonical geometries [Figs.~\ref{fig:3}(a,d)]. On log--log axes, this curve has slope $1/2$. In the linear representation [Figs.~\ref{fig:3}(b,e)], the data approach $|\Shnd|\approx\sqrt{\wstar}$, in agreement with Eq.~\eqref{eq:B3}. At $\wstar=10^4$, $|\Shnd|/\sqrt{\wstar}$ ranges from 0.99 to 1.06 across all geometries and flow angles, while the phase ranges from $41.8^\circ$ to $44.3^\circ$. For each of G-1 through G-5, the difference in $|\Shnd|$ between flow angles is at most 0.003 at this frequency. Thus, both the exponent and the leading prefactor are universal.

The crossover to the high-frequency limit depends on the pseudo-steady transfer rate. The magnitude $|\Shnd|$ remains near its low-frequency plateau $\avg{Sh}$ until it approaches the high-frequency asymptote $|\Shnd|\approx\sqrt{\wstar}$. Equating these two expressions gives the crossover frequency
$\wstar_c\sim\avg{Sh}^{\,2}$. Geometries with larger pseudo-steady transfer rates therefore leave the plateau at higher frequencies. For G-6, $\wstar_c\approx760$; at $\wstar=10^3$, its phase is only $24.6^\circ$ and $|\Shnd|/\sqrt{\wstar}=1.18$. Thus, the high-frequency limit cannot be assessed at one fixed frequency for all geometries. The diffusion length must also be smaller than the narrowest local pore dimension, which imposes an additional geometric constraint.

At large $\Pe$, the phase response can become nonmonotonic [Fig.~\ref{fig:3}(c), G-2 at $\Pe=1000$] because of competition between advective redistribution and transverse diffusion. This competition affects the crossover and intermediate-frequency response but does not alter the high-frequency exponent or leading prefactor.

\subsection{Geometric correction}

\begin{figure}[ht!]
    \includegraphics[width=0.98\linewidth]{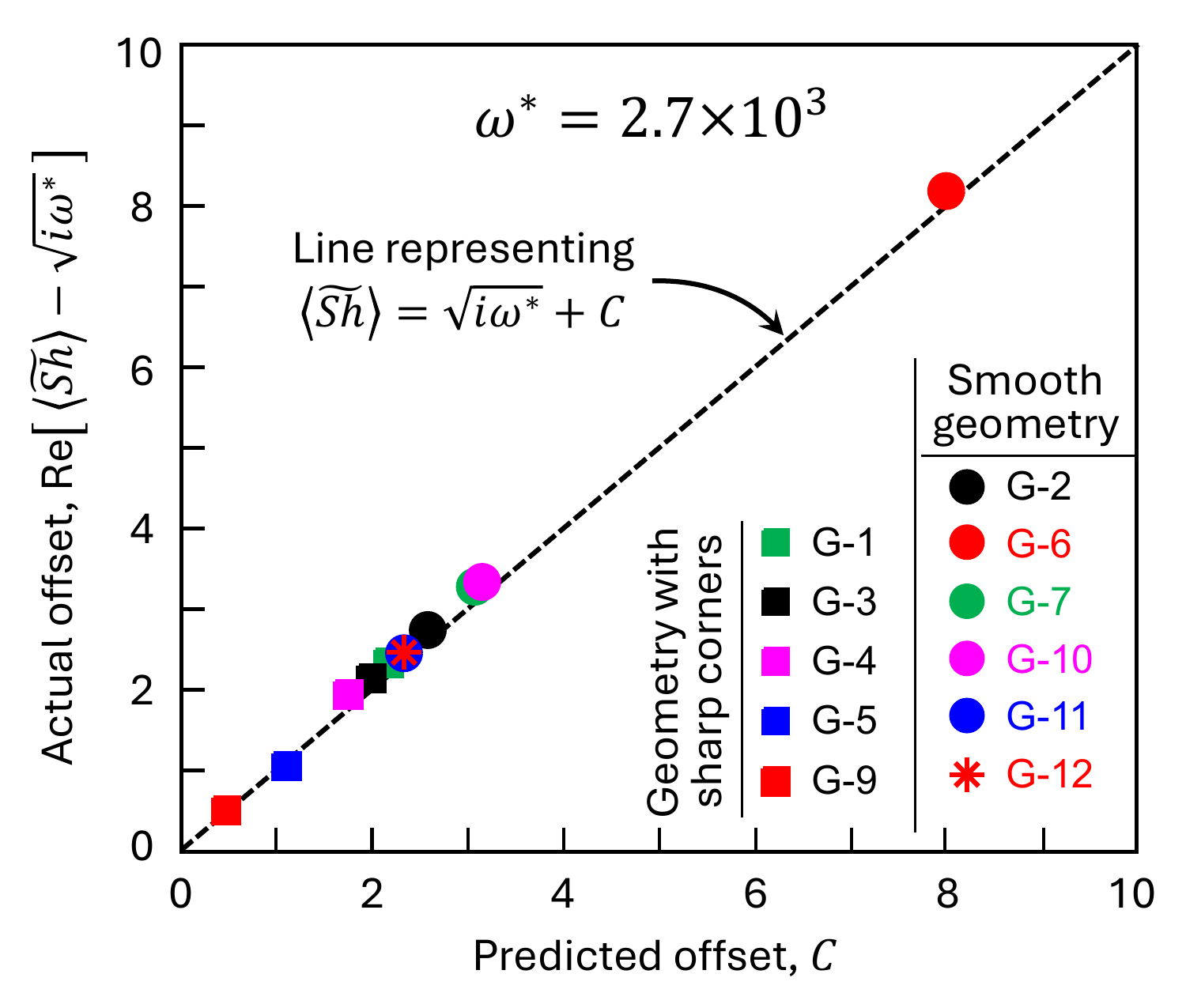}
    \caption{First correction to the Warburg limit for $\Da\to\infty$ and $\Pe=100$. The measured offset, $\mathrm{Re}[\Shnd-\sqrt{\iu\wstar}]$ at $\wstar\approx2.7\times10^3$, is plotted against the predicted offset $C$ from Eq.~\eqref{eq:C_general} for the geometries listed in Table~\ref{tab:2}. Circles denote smooth inclusions, for which the geometric contribution is determined by curvature, and squares denote polygonal inclusions, for which it is determined by the edge sum. G-12 is marked by a star because it would otherwise overlap with G-11. The dashed line indicates perfect agreement. G-8 is omitted because its two-arc boundary was not evaluated.}
    \label{fig:4}
\end{figure}

\begin{table}[t!]
    \centering
    \caption{Predicted and measured high-frequency offsets for $\Da\to\infty$, $\Pe=100$, and $\lP=d_P$. The storage term is $\sA\lP/\eps=4(1-\eps)/\eps$. The geometric term is $\pi d_P/\Atot$ for smooth rods and $-(\lP/\Atot)\sum_e \Lambda(\gamma_e)$ for polygons, from Eq.~\eqref{eq:C_general}. The predicted offset $C$ is their sum. The measured offset is $\Shnd-\sqrt{\iu\wstar}$ at $\wstar\approx2.7\times10^3$; its real part is compared with $C$, and its imaginary part, which vanishes in the asymptotic limit, indicates how far each case is from convergence. Porosities are computed from the exact shapes. G-8 is omitted because its two-arc boundary was not evaluated.}
    \label{tab:2}
    \begin{tabular*}{\columnwidth}{@{\extracolsep{\fill}}lccccc}
        \toprule
        & & & & \multicolumn{2}{c}{Measured} \\
        \cmidrule(l){5-6}
        ID & Storage & Geometric & $C$ & Re & Im \\
        \midrule
        G-1  & 1.56 & 0.60    & 2.16 & 2.32 & $-0.13$ \\
        G-3  & 1.56 & 0.43    & 1.99 & 2.13 & 0.16    \\
        G-4  & 1.56 & 0.19    & 1.75 & 1.93 & $-0.29$ \\
        G-5  & 1.56 & $-0.45$ & 1.11 & 1.04 & $-0.43$ \\
        G-9  & 0.35 & 0.12    & 0.47 & 0.49 & $-0.77$ \\
        \midrule
        G-2  & 1.58 & 1.00    & 2.58 & 2.73 & 0.23    \\
        G-6  & 6.99 & 1.00    & 7.99 & 8.18 & $-1.45$ \\
        G-7  & 2.19 & 0.88    & 3.07 & 3.27 & 0.18    \\
        G-10 & 2.15 & 1.00    & 3.15 & 3.33 & 0.16    \\
        G-11 & 1.33 & 1.00    & 2.33 & 2.44 & 0.22    \\
        G-12 & 1.33 & 1.00    & 2.33 & 2.45 & 0.22    \\
        \bottomrule
    \end{tabular*}
\end{table}

Once the conditions in Eq.~\eqref{eq:C_conditions} are satisfied, Eq.~\eqref{eq:C_general} predicts that $\Shnd-\sqrt{\iu\wstar}$ approaches a real, geometry-dependent constant. We therefore compare $C$ with the real part of this offset. This choice follows from the expansion
\begin{equation}
    \Shnd-\sqrt{\iu\wstar}=C+\frac{C_1}{\sqrt{\iu\wstar}}+O\!\left(\frac{\Pe}{\wstar}\right)+\cdots,
    \label{eq:offset_expansion}
\end{equation}
where $C_1$ is a geometric coefficient associated with the next-order correction. The constant $C$ is real because it depends only on the real quantities $\sA$, $\eps$, $\lP$, $\avg{\rho}_s$, and $\Lambda(\gamma)$. In the derivation, the storage contribution has the form $a_\infty\approx-\sA/(\eps m)$, where $m=\sqrt{\iu\omega/D}$. Multiplication by the leading wall-normal gradient $m$ cancels the complex factor, producing a real contribution to $C$. Thus, $C$ represents an additional transfer response in phase with the concentration difference. The higher-order terms in Eq.~\eqref{eq:offset_expansion} are complex and vanish as $\wstar$ increases. Consequently, the real part of the measured offset converges to $C$, while its imaginary part measures the remaining deviation from the asymptotic limit.

The magnitude $|\Shnd-\sqrt{\iu\wstar}|$ provides a less reliable estimate of $C$. If $\mathcal{R}$ denotes the higher-order remainder, then
$|C+\mathcal{R}|\approx C+\mathrm{Re}\,\mathcal{R}+(\mathrm{Im}\,\mathcal{R})^2/(2C)$. The last term is positive and becomes largest when $C$ is small. The magnitude is therefore biased upward when the solution has not fully converged. For G-9, for example, $C=0.47$, while the imaginary part of the offset at $\wstar\approx2.7\times10^3$ is $-0.77$. The magnitude is then 0.92, nearly twice the real part, which is 0.49. For most geometries, the imaginary part has magnitude no greater than 0.3 at this frequency, and the real part and magnitude differ by less than 0.03. The exceptions are G-5, G-6, and G-9, with imaginary parts of $-0.43$, $-1.45$, and $-0.77$, respectively. Their real parts should therefore be interpreted with caution because the residual contains several contributions, including the next geometric correction, advection of order $\Pe/\wstar$, and numerical error. These contributions cannot be separated at a single frequency.

The offsets are evaluated at $\wstar\approx2.7\times10^3$. This frequency satisfies Eq.~\eqref{eq:C_conditions} for every geometry. It is above the crossover frequency $\wstar_c\sim\avg{Sh}^2$, which is at most approximately 760 for G-6, gives $\Pe/\wstar\approx0.04$ at $\Pe=100$, and lies within the range over which the numerical solutions are grid-converged.

Table~\ref{tab:2} and Fig.~\ref{fig:4} compare the measured offsets with the theoretical predictions. For the six smooth rods, the predicted offsets range from approximately 2.4 to 8.2, while the measured values are higher by approximately 0.1--0.2. G-6 has the largest offset because it has the lowest porosity. Its slow approach to the Warburg limit in Fig.~\ref{fig:3} is therefore a storage effect rather than a departure from the universal scaling. For G-6 at $\wstar=10^4$, Eq.~\eqref{eq:approach} gives a phase of $41.8^\circ$ and a magnitude ratio of 1.06, consistent with the slow approach shown in Fig.~\ref{fig:3}. Subtracting $C$ reduces the G-6 residual
$|\Shnd-\sqrt{\iu\wstar}-C|/\sqrt{\wstar}$ from 0.086 to 0.006 at this frequency.

The edge term in Eqs.~\eqref{eq:C_general} and \eqref{eq:Lambda_gamma} extends this agreement to polygonal inclusions. G-1, G-3, and G-4 lie approximately 0.15 above the prediction. G-5 is a staircase with 13 convex and 9 reentrant corners. The reentrant corners make its geometric contribution negative, so its measured offset lies below the storage contribution alone. G-5 therefore provides the most discriminating test of the edge term. Treating its boundary as smooth and using Eq.~\eqref{eq:C_2D} gives 1.95, compared with the measured value of 1.04 and the predicted value of 1.11. For G-1, G-3, and G-4, the smooth- and sharp-boundary estimates differ by less than 0.2. Their measured offsets show the same small positive bias as the smooth rods, so these geometries do not by themselves distinguish between the two limits. G-11 and G-12 differ only in the orientation of the RVE. Their plateaus differ substantially, 9.1 and 5.3, whereas their offsets agree closely, 2.44 and 2.45. This agreement occurs because the storage and curvature contributions do not depend on the array orientation.

The conditions in Eq.~\eqref{eq:C_conditions} are also supported by the data. At $\Pe=0.1$, $10$, and $100$, the G-2 and G-4 offsets at $\wstar=10^3$ agree within approximately 0.1. At $\Pe=1000$, the offsets collapse only near $\wstar\approx7\times10^3$, where $\Pe/\wstar\approx0.15$. This behavior is consistent with $\wstar\gg\Pe$ and not with the weaker condition $\wstar\gg\Pe^{2/3}\approx100$. The results for different flow angles and array orientations likewise confirm that $C$ is independent of flow direction.

\subsection{Impedance}

\begin{figure}[t]
    \includegraphics[width=0.95\linewidth]{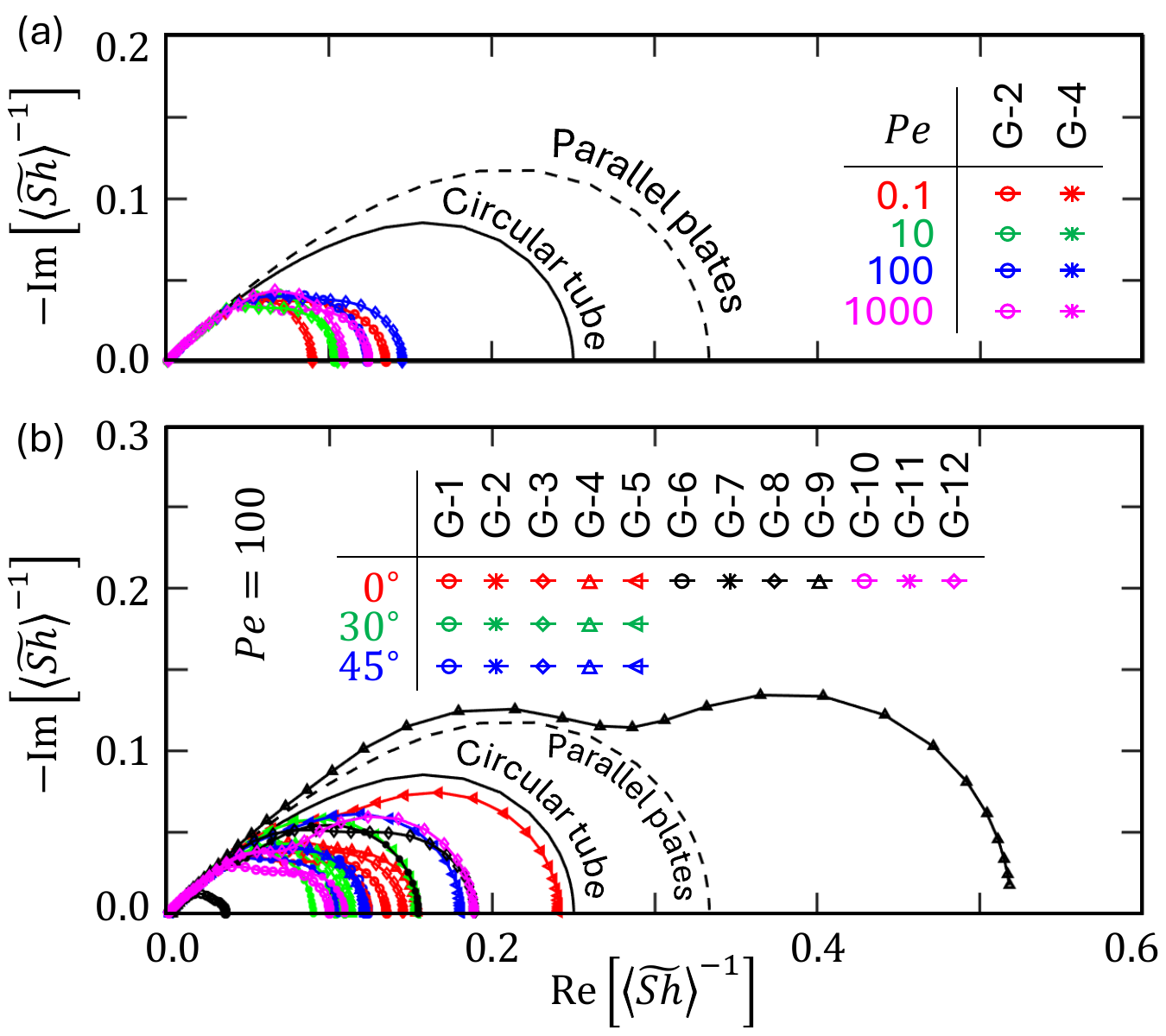}
    \caption{Nyquist plots of $1/\Shnd$ for $\Da\to\infty$. (a)~G-2 and G-4 at $\theta=0^\circ$ for $\Pe=0.1$, $10$, $100$, and $1000$. (b)~All twelve microstructures at $\Pe=100$. In (b), red, green, and blue denote $\theta=0^\circ$, $30^\circ$, and $45^\circ$ for G-1 to G-5, respectively; black denotes G-6 to G-9; and magenta denotes G-10 to G-12, all at $\theta=0^\circ$. The dashed and solid black curves denote the parallel-plate and circular-tube solutions, respectively.}
    \label{fig:5}
\end{figure}

Figure~\ref{fig:5} presents Nyquist plots of ${\Shnd}^{-1}$. According to Eq.~\eqref{eq:ZSh}, this quantity differs from $\Znd$ only by the capacitive term. It therefore represents the diffusion-controlled contribution to the impedance in the facile-kinetics limit, $\Da\to\infty$.

At high frequency, all curves approach the same unit-slope Warburg response. At lower frequencies, they separate into geometry-dependent arcs that close on the real axis at $1/\avg{Sh}$. For G-2 and G-4, changing $\Pe$ alters the arc size but not the high-frequency response [Fig.~\ref{fig:5}(a)]. For the twelve microstructures at $\Pe=100$ [Fig.~\ref{fig:5}(b)], the real-axis intercepts range from approximately 0.04 for G-6 to approximately 0.52 for G-9. Thus, the high-frequency impedance samples the local interface, whereas the low-frequency response reflects the global pore geometry~\cite{Caprani1987}.

G-9 is the only geometry whose spectrum contains two lobes. A single lobe typically reflects a dominant diffusion time across the pore. The thin inclined strip in G-9 creates two distinct transport distances. Fluid near the long faces of the strip responds at moderate frequencies, whereas fluid in the corners of the unit cell beyond the strip ends responds only at lower frequencies. The second lobe is consistent with this slower transport process. The phase response of G-9 in Fig.~\ref{fig:3}(f) supports this interpretation, departing from zero at a lower frequency than for any other geometry. At high frequency, however, G-9 joins the common Warburg response. Thus, a unit cell with widely separated pore dimensions can produce multiple spectral lobes. This possibility should be considered when separate features in measured impedance spectra are attributed to distinct physical processes.

\subsection{Comparison with experiments}
\label{sec:experiments}

\begin{figure*}[ht!]
    \includegraphics[width=0.85\linewidth]{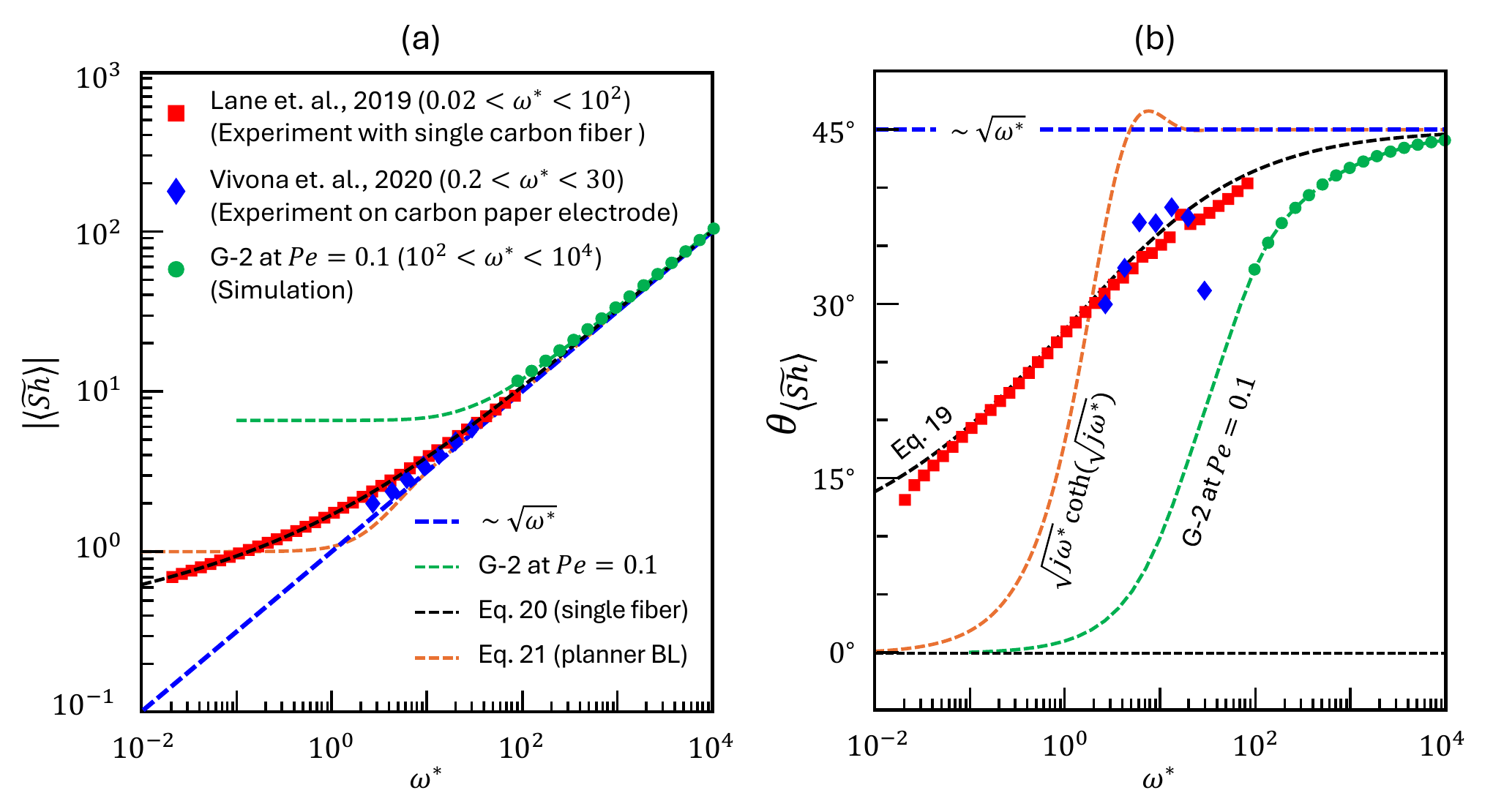}
    \caption{Comparison of the simulated and measured spectral Sherwood number with the single-fiber and planar-boundary-layer limits for $\Da\to\infty$. (a)~Magnitude and (b)~phase of $\Shnd$. The green dashed curve is G-2 at $\Pe=0.1$, with $\lP=d_P=d$, and green circles mark its points at $\wstar\geq100$. Red squares are the impedance data of Ref.~\cite{LandonLane2019} for a single carbon-felt fiber in ferro/ferricyanide at the reversible potential, converted with Eqs.~\eqref{eq:LL_ZF}--\eqref{eq:LL_conversion} and $\lP=d$. Blue diamonds are the spectrum of Ref.~\cite{Vivona2020} for a vanadium symmetric cell at 0.05~A\,cm$^{-2}$ and 20~ml\,min$^{-1}$, converted with Eq.~\eqref{eq:V_conversion} and $\lP=d_{BL}=20~\mu$m. Only the band from 0.1 to 1.1~Hz, attributed to transport from the pore center to the fiber surface, is shown. Their magnitudes are plotted as $\Shnd/\Da_{\mathrm{eff}}$, that is, for $\Da_{\mathrm{eff}}=1$, whereas their phase is independent of $\Da_{\mathrm{eff}}$. The black dashed curve is the single-fiber solution, Eq.~\eqref{eq:Sh_fiber}, the orange dashed curve is the planar boundary layer, Eq.~\eqref{eq:Sh_planar}, and the blue dashed lines mark the Warburg limit, $|\Shnd|=\sqrt{\wstar}$ and $\theta_{\Shnd}=45^\circ$.}
    \label{fig:exp}
\end{figure*}

Figure~\ref{fig:exp} compares the theory and one of the present simulations with two published impedance measurements. The experiments probe different geometries and use different characteristic lengths. We therefore first describe the reference solutions and then explain how each data set is converted to $\Shnd$ and $\wstar$.

Two limiting solutions correspond to the experimental geometries. For a single cylinder of diameter $d$ in an unbounded fluid without flow, Eq.~\eqref{eq:a_closure} reduces to radial diffusion with $a_\infty\to0$, because the fluid volume is unbounded and the zero-mean constraint does not affect the solution near the fiber. With $\lP=d$, Eq.~\eqref{eq:Sh} gives
\begin{equation}
    \Shnd=x\,\frac{K_1(x/2)}{K_0(x/2)},\qquad x=\sqrt{\iu\wstar},
    \label{eq:Sh_fiber}
\end{equation}
where $K_0$ and $K_1$ are modified Bessel functions of the second kind. This result agrees with the known Bessel-function impedance of a cylindrical electrode~\cite{Jacobsen1995,LandonLane2019}. At high frequency, it approaches $x+1$, consistent with Eq.~\eqref{eq:C_general}, because the storage term vanishes in an unbounded fluid and $\lP\avg{\rho}_s=1$ for a cylinder with $\lP=d$. At low frequency, $|\Shnd|$ decreases without bound and its phase remains finite because a two-dimensional diffusion field around an isolated cylinder has no steady state.

For a planar layer of thickness $\lP$ bounded by a reactive surface on one side and a fixed-concentration boundary on the other,
\begin{equation}
    \Shnd=x\coth x,\qquad x=\sqrt{\iu\wstar},
    \label{eq:Sh_planar}
\end{equation}
which approaches 1 at low frequency and $x$ at high frequency.

We next convert the single-fiber data of Landon-Lane \textit{et al.}, who measured the impedance of an isolated carbon-felt fiber in 10~mM each of ferrocyanide and ferricyanide in 1~M KNO$_3$, at the reversible potential and without forced flow~\cite{LandonLane2019}. They fitted the spectrum with a series resistance $R_u$, a constant-phase element (CPE) with parameters $Y_0$ and $\phi$ for the double layer, and a faradaic impedance
\begin{equation}
    \begin{aligned}
        &Z_F=R_{ct}\left[1+\psi\sqrt{\tau_d}\,W(\xi)\right],\\
        &W(\xi)=\frac{K_0(\xi)}{\xi\,K_1(\xi)},\qquad \xi=\sqrt{\iu\omega\tau_d},
    \end{aligned}
    \label{eq:LL_ZF}
\end{equation}
where $R_{ct}$ is the charge-transfer resistance, $\tau_d=r_0^2/D$ is the diffusion time, $r_0$ is the fiber radius, and $\psi=(k_f+k_b)/\sqrt{D}$, with $k_f$ and $k_b$ denoting the forward and backward rate constants. At the reversible potential with equal concentrations of the two species, $k_f=k_b=k^0$, so $\psi=2k^0/\sqrt{D}$. The fitted values are $R_u=790~\Omega$, $Y_0=56$~nF\,s$^{\phi-1}$, $\phi=0.843$, $R_{ct}=6470~\Omega$, $\psi=5.28$~s$^{-1/2}$, and $\tau_d=13.7$~ms~\cite{LandonLane2019}.

We convert the measured impedance $Z(\omega)$ in three steps. First, we remove the series resistance and the double-layer branch to isolate the faradaic impedance,
\begin{equation}
    Z_F=\left[\frac{1}{Z-R_u}-Y_0(\iu\omega)^{\phi}\right]^{-1}.
    \label{eq:LL_remove}
\end{equation}
Second, we remove the charge-transfer contribution by normalizing with $R_{ct}$ and subtracting 1, which leaves the mass-transfer contribution,
$Z_F/R_{ct}-1=\psi\sqrt{\tau_d}\,W(\xi)$. Third, we match the resulting expression to the dimensionless variables of the closure problem. We use $\lP=d=2r_0$, as for the circular rods, so that
\begin{equation}
    \wstar=\frac{\omega d^2}{D}=4\omega\tau_d,\qquad \xi=\frac{x}{2},\qquad W=\frac{2}{\Shnd}.
    \label{eq:LL_scaling}
\end{equation}
The last relation follows by comparing $W(\xi)$ with Eq.~\eqref{eq:Sh_fiber}. Combining these steps gives
\begin{equation}
    \frac{1}{\Shnd}=\frac{Z_F}{R_{ct}\Da_c}-\frac{1}{\Da_c},\qquad\Da_c=2\psi\sqrt{\tau_d}=\frac{2k^0d}{D}.
    \label{eq:LL_conversion}
\end{equation}
This expression has the same form as the impedance of the closure problem with finite kinetics. The term $1/\Da_c$ is the frequency-independent kinetic resistance, while $1/\Shnd$ is the mass-transfer resistance. The subscript $c$ indicates that $\Da_c$ refers to the redox couple. Because both species contribute to the diffusion impedance, the effective rate constant is twice $k^0$, and the kinetic crossover occurs at $\omega=4(k^0)^2/D$ rather than at $(k^0)^2/D$.

This conversion requires only the fitted values of $\tau_d$ and $\psi$ and the ratio $Z_F/R_{ct}$. It therefore requires neither the electrode area nor the fiber length. The diffusivity also cancels from both $\wstar=4\omega\tau_d$ and $\Da_c=2\psi\sqrt{\tau_d}$. For reference, the authors' diffusivity, $D=8.5\times10^{-10}$~m$^2$/s, obtained as the mean of the values for the two species from limiting-current measurements, corresponds to $r_0=\sqrt{D\tau_d}=3.42~\mu$m and $\lP=d=6.84~\mu$m~\cite{LandonLane2019}. The resulting values are $\wstar=0.344f$, with $f$ in Hz, and $\Da_c=1.24$. The published spectrum spans 0.01~Hz to 243~kHz, or $\wstar\approx0.003$ to $8\times10^4$. We use the range $\wstar=0.021$ to 84, corresponding to $f=0.062$ to 245~Hz. Below this range, the authors report a low-frequency departure that may result from convection. Above it, $Z_F/R_{ct}-1$ is small compared with the uncertainty in $R_{ct}$, so subtracting the kinetic term is dominated by noise.

We next convert the flow-battery data. Vivona \textit{et al.}\ measured symmetric cells with the negative vanadium electrolyte and carbon-paper electrodes with a thickness of 230~$\mu$m~\cite{Vivona2020}. We use their spectrum at 0.05~A\,cm$^{-2}$ and 20~ml\,min$^{-1}$. Their model describes transport from the pore center to the fiber surface as diffusion across a planar boundary layer with assumed thickness $d_{BL}=20~\mu$m. The fitted effective diffusivity is
$D_{BL}=4.8\times10^{-4}d_{BL}$~cm$^2$\,s$^{-1}$, or $9.6\times10^{-11}$~m$^2$/s. This effective diffusivity includes both molecular diffusion and convection within the pores. The corresponding impedance element is $\tanh(x)/x$, where $x=d_{BL}\sqrt{\iu\omega/D_{BL}}$~\cite{Vivona2020}. Setting $\lP=d_{BL}$ and replacing $D$ with $D_{BL}$ maps this element onto Eq.~\eqref{eq:Sh_planar}, with
\begin{equation}
    \wstar=\frac{\omega d_{BL}^2}{D_{BL}}=2\pi f\times4.17~\mathrm{s},\qquad x=\sqrt{\iu\wstar}.
    \label{eq:V_scaling}
\end{equation}
Thus, the characteristic length and diffusivity for this data set come from the authors' pore-scale model rather than from a measured particle size and molecular diffusivity.

The spectrum contains a charge-transfer arc with a peak near 1.4~kHz, a low-frequency arc with a peak near 0.03~Hz attributed to transport through the electrode thickness, and a linear feature between them, from approximately 0.1 to 10~Hz, attributed to transport from the pore center to the fiber surface~\cite{Vivona2020}. Because the charge-transfer arc peaks near 1.4~kHz, the double-layer current is negligible below approximately 1~Hz, where $\omega R_{ct}C_{\mathrm{dl}}<10^{-2}$. In this range, the measured impedance reduces to $Z\approx R_\Omega+R_{ct}+Z_W$, where $R_\Omega$ is the ohmic resistance and $Z_W$ is the mass-transfer impedance. We obtain $R_\Omega=0.49$ and $R_{HF}=R_\Omega+R_{ct}=1.28~\Omega$\,cm$^2$ as the two real-axis intercepts of a circle fitted to the charge-transfer arc, giving $R_{ct}=0.79~\Omega$\,cm$^2$. We then remove the kinetic contribution by subtracting $R_{HF}$ and normalize the mass-transfer contribution by $R_{ct}$:
\begin{equation}
    \frac{\Shnd}{\Da_{\mathrm{eff}}}=\frac{R_{ct}}{Z-R_{HF}}=\frac{R_{ct}}{Z_W}.
    \label{eq:V_conversion}
\end{equation}
This ratio is independent of the electrode area. The factor of 2 associated with the two electrodes in the symmetric cell also cancels.

The effective Damk\"ohler number $\Da_{\mathrm{eff}}$ cannot be determined from these data alone. In the authors' model, it equals the sum of the forward and backward reaction rates at the operating point multiplied by $d_{BL}/D_{BL}$. Their fitted rate constant, $k_f=1.4\times10^{-4}$~m/s, gives $k_fd_{BL}/D_{BL}\approx29$, but this value excludes the overpotential-dependent factors in the rate expression. If $\Da_{\mathrm{eff}}=29$, the converted magnitudes would lie approximately 30 times above both reference curves. Conversely, the measured ratio $R_{ct}/Z_W$ follows Eq.~\eqref{eq:Sh_planar} for $\Da_{\mathrm{eff}}\approx0.7$--$0.9$. We therefore plot the magnitude for $\Da_{\mathrm{eff}}=1$ and use the phase for the quantitative comparison, since it is independent of $\Da_{\mathrm{eff}}$.

The usable frequency range is limited at both ends. Below approximately 0.1~Hz, transport through the electrode thickness dominates, so the response no longer represents pore-scale transfer. Above approximately 1~Hz, $Z_W$ becomes small compared with $R_{ct}$, its value is comparable to the uncertainty in $R_{HF}$, and the published data points overlap. We therefore use the range 0.1 to 1.1~Hz, corresponding to $\wstar\approx3$ to 30.

Among the present simulations, G-2 at $\Pe=0.1$ is the closest counterpart to these experiments for three reasons. First, G-2 is a circular rod, so with $\lP=d_P=d$ it has the same characteristic length and curvature contribution, $\lP\avg{\rho}_s=1$, as the single fiber in Eq.~\eqref{eq:Sh_fiber}. Second, $\Pe=0.1$ is the weakest simulated flow and therefore most closely matches the quiescent fiber experiment. At this P\'eclet number, the advective contribution of order $\Pe/\wstar$ is negligible over the plotted frequency range, and even the pseudo-steady plateau is essentially that of pure diffusion. Third, G-2 has a relatively high porosity, $\eps=0.717$, so its storage contribution, $4(1-\eps)/\eps=1.58$, is moderate. G-11 and G-12 have slightly higher porosity, 0.751, but were simulated only at $\Pe=100$ and in rectangular unit cells. G-6, with $\eps=0.364$ and a storage contribution of 7.0, is the least suitable comparison, while the polygonal inclusions do not share the curvature contribution of a fiber.

The single-fiber data follow Eq.~\eqref{eq:Sh_fiber} in both magnitude and phase over the full range shown. At $\wstar\approx34$, for example, $|\Shnd|=6.4$ (5.9 to 6.8), compared with the predicted value of 6.5, and the phase is $37.9^\circ$, compared with the predicted $39.4^\circ$. The offset $\mathrm{Re}[\Shnd-\sqrt{\iu\wstar}]$, averaged over $\wstar=5$ to 35, is $0.94\pm0.13$, consistent with the exact value of approximately 0.9 over this range and with the asymptotic value $C=1$. This comparison nevertheless has several limitations. The parameters were fitted by the original authors using a cylindrical diffusion model, so part of the agreement in shape is built into the fit, although $C$ was not fitted directly. The fitted radius of 3.42~$\mu$m is also smaller than the mean radius of approximately 5~$\mu$m measured by microscopy~\cite{LandonLane2019}. Finally, an isolated fiber tests only the curvature contribution, not the storage contribution in Eq.~\eqref{eq:C_general}.

At low frequency, G-2 lies on a plateau, $|\Shnd|\approx6.6$, with a phase close to zero, whereas the fiber data continue to decrease in magnitude and retain a phase of $13^\circ$ to $40^\circ$. This difference arises from the geometry rather than from the transport law. G-2 is a periodic array in which neighboring rods are separated by approximately $0.67d$. When the diffusion length $\delta=d/\sqrt{\wstar}$ exceeds this gap, the concentration disturbance fills the pore. The zero-mean constraint in Eq.~\eqref{eq:a_closure} then fixes $a_\infty$, and the response becomes pseudo-steady, with a finite $\avg{Sh}$ and a vanishing phase. An isolated fiber has no neighboring surface or bounded pore. Its diffusion layer therefore continues to grow as the frequency decreases, and no pseudo-steady plateau is reached. The two geometries can agree only when the diffusion layer is thin compared with the pore, that is, above the crossover $\wstar_c\sim\avg{Sh}^2\approx44$ for G-2. Above a few times this frequency, they differ mainly through the storage contribution, as predicted by Eq.~\eqref{eq:C_general}. After subtracting $4(1-\eps)/\eps=1.58$, $|\Shnd|$ for G-2 agrees with Eq.~\eqref{eq:Sh_fiber} to within 0.3\% from $\wstar\approx200$ to $10^3$. The phase of G-2 remains lower than that of the fiber by $2^\circ$ at $\wstar=10^3$ and by $1.1^\circ$ at $\wstar=2.7\times10^3$, consistent with the difference $1.58/\sqrt{2\wstar}$ predicted by Eq.~\eqref{eq:approach}. The fiber measurements, however, end at $\wstar\approx90$, where the two geometries necessarily differ.

Over the range 0.1 to 1.1~Hz, the phase of the converted Vivona data lies between $30^\circ$ and $38^\circ$, below both the planar boundary-layer prediction of Eq.~\eqref{eq:Sh_planar}, which ranges from $37^\circ$ to $47^\circ$, and the Warburg limit of $45^\circ$. The planar phase briefly exceeds $45^\circ$, reaching a maximum of $46.6^\circ$ near $\wstar\approx8$, because the concentration disturbance reflected from the fixed-concentration boundary returns to the interface with a phase lag. This overshoot is a property of the finite planar layer and does not occur in the Warburg limit. A phase below $45^\circ$ is expected for a complete porous electrode. Transport is distributed through the electrode thickness, and the low-frequency arc associated with through-plane transport overlaps the lower end of the frequency band. Both effects reduce the phase relative to that of a single interface. The single-fiber phase over the same range of $\wstar$ is $32^\circ$ to $39^\circ$. The proximity of these points to the single-fiber curve in Fig.~\ref{fig:exp} is partly coincidental because the two data sets use different characteristic lengths, $d_{BL}$ and $d$. The inferred $\Da_{\mathrm{eff}}$ of order 1 also has a direct interpretation. At 1~Hz, $Z_W$ is approximately 0.13~$\Omega$\,cm$^2$, compared with $R_{ct}=0.79~\Omega$\,cm$^2$, so the electrode is mainly kinetic-limited in this frequency band. Within the band, the phase is lowest at both ends, $30^\circ$ at $\wstar\approx3$ and $31^\circ$ at $\wstar\approx30$, and reaches $38^\circ$ near $\wstar\approx13$. The lower end overlaps the through-plane arc, while at the upper end $Z_W$ becomes comparable to the uncertainty in $R_{HF}$. The central values are therefore the most representative of pore-scale transfer. This result agrees with the conclusion of Sec.~\ref{sec:applications} that vanadium electrodes lie above the kinetic limit of the operating window, $\wstar\ll\Da^2$.

These comparisons are consistent with the predicted transfer law and with the curvature contribution to the first correction. They do not provide an independent validation of the porous-medium simulations because the experiments do not reach the frequencies at which G-2 follows the high-frequency limit, and because the conversion parameters come from the original authors' model fits. A direct test would require spectra measured for the same microstructure or a simulation of an isolated circle in a large unit cell. Such a system should follow Eq.~\eqref{eq:Sh_fiber} over the full experimental frequency range.

Direct pulse measurements provide an additional comparison. In our earlier work, square-wave current pulses applied to graphite-foil electrodes with channels approximately 100~$\mu$m wide in 15~mM ferro/ferricyanide produced current densities above the classical film-theory prediction. The measurements agreed with a theory containing no adjustable parameters~\cite{Hamid2023}. These experiments support the transient limits developed below. To our knowledge, however, the frequency-dependent $\Shnd$ and the storage contribution to $C$ have not yet been measured directly in a porous electrode.
    \section{Implications for electrode operation}

\subsection{What high-frequency transfer changes}

High-frequency modulation can produce interfacial fluxes that exceed the steady limiting flux because the concentration disturbance penetrates only a thin region near the interface. It does not increase the steady limiting flux. Because the transport problem is linear, the time-averaged flux depends only on the zero-frequency response,
$j_{\mathrm{av}}=(D/\lP)\avg{Sh}\,\Delta c_{\mathrm{av}}$, where $\Delta c_{\mathrm{av}}$ is the time average of $c_{s,eq}-\avg{c}$. Since the interfacial concentration cannot be negative, the time-averaged flux to a consuming electrode satisfies
\begin{equation}
    |j_{\mathrm{av}}|\leq j_{\mathrm{lim}}=\frac{D}{\lP}\avg{Sh}\avg{c}.
    \label{eq:average_bound}
\end{equation}
This is the porous-electrode form of the classical bound for pulse electrolysis~\cite{Cheh1971,Ibl1980}. High-frequency transfer therefore differs fundamentally from the steady OLC mechanisms discussed above, which increase the time-averaged current through electrokinetic effects or additional charge carriers~\cite{Dydek2011,Nam2015}. Its benefit is instead an increase in the peak flux. It requires neither surface charge nor confinement and is therefore complementary to these mechanisms. It also differs from morphology-related improvements in pulse-interrupt electrodeposition~\cite{Zelger2016}, which arise from surface diffusion during the pulse-off period.

Consider a sinusoidal interfacial concentration that reaches zero at the end of each cycle, while $\avg{c}$ remains fixed. The peak flux relative to the steady limiting flux is then
\begin{equation}
    \frac{j_{\mathrm{peak}}}{j_{\mathrm{lim}}}=1+\frac{|\Shnd|}{\avg{Sh}}.
    \label{eq:peak}
\end{equation}
At high frequency, the microstructure affects this ratio only through $\avg{Sh}$. Consequently, geometries with smaller pseudo-steady Sherwood numbers exhibit the largest relative enhancement.

\subsection{Pulse duration}

Consider a pulse with flux $j=Jj_{\mathrm{lim}}$, where $J\gg1$. In the thin-layer regime, the interfacial concentration reaches zero at the Sand transition time~\cite{BardFaulkner2001},
\begin{equation}
    \tau^*=\frac{\tau D}{\lP^2}=\frac{\pi}{4J^2\avg{Sh}^2}.
    \label{eq:Sand}
\end{equation}
Equation~\eqref{eq:Sand} relates the maximum pulse duration to the same pseudo-steady Sherwood number that determines the crossover frequency. For example, G-2 at $\Pe=100$ has $\avg{Sh}\approx8.0$ and $d_P=15~\mu$m. With $D=7\times10^{-10}$~m$^2$/s, a pulse with $J=10$ can be sustained for approximately $40~\mu$s.

\subsection{Pulse length in a flow-battery electrolyte}

In dimensional form, the transition time is
$\tau=\pi D\avg{c}^2/(4j^2)$, so it increases with the square of the concentration at fixed flux. Consider a 100~mM TEMPO electrolyte with $D=2\times10^{-9}$~m$^2$/s~\cite{Stumme2023}, a pore length of 10~$\mu$m, and $\avg{Sh}\approx8$. A flux corresponding to 1~A/cm$^2$ of interfacial area is approximately 6.5 times the steady limiting flux and can be sustained for about 1.5~ms.

Increasing the concentration can therefore extend the usable pulse duration, but only within practical limits. For TEMPO in acetonitrile, both $D$ and $k^0$ decrease at high concentrations of the active species and supporting salt, while the conductivity approaches a plateau near 1~M TBAPF$_6$ (tetrabutylammonium hexafluorophosphate)~\cite{Stumme2023}. Concentrated electrolytes therefore trade longer pulse durations for slower reaction kinetics and mass transport. These measured concentration-dependent properties can be incorporated directly into Eqs.~\eqref{eq:Sand} and \eqref{eq:window_dim}.

\subsection{Operating window}

Combining the conditions above with the finite-kinetics saturation gives the frequency window in which
$\Shnd\approx\sqrt{\iu\wstar}+C$ applies:
\begin{equation}
    \max\left[\avg{Sh}^2,\;(\lP/g)^2,\;\Pe\right]\ll\wstar\ll\Da^2.
    \label{eq:window}
\end{equation}
For $\lP=10~\mu$m and $D=7\times10^{-10}$~m$^2$/s, $\wstar=10^3$ corresponds to a frequency of approximately 1~kHz. The upper bound then requires $\Da\gg30$. In dimensional form, this upper bound is
\begin{equation}
    \omega\ll\frac{k^2}{D},
    \label{eq:window_dim}
\end{equation}
which is independent of the pore length. Increasing the pore length raises $\Da$ and lowers the frequency required to reach the high-frequency regime, but it does not increase the upper frequency limit. Thus, the ratio $k^2/D$, rather than $\Da$ alone, determines whether a given chemistry can access the Warburg regime. We define the corresponding kinetic frequency as
$f_k=k^2/(2\pi D)$.

The reaction-rate constant $k$ in Eq.~\eqref{eq:pore2} is an effective linearized constant. Its value can increase with overpotential and may exceed its equilibrium value by approximately a factor of 100 at an overpotential of 0.24~V for a transfer coefficient of 0.5. The linear analysis is therefore only approximate under strongly polarized conditions.

\begin{table*}[t!]
    \centering
    \caption{Reported standard rate constants and the resulting kinetic limits. $\Da$ is evaluated at $\lP=10~\mu$m with $k=k^0$, and $f_k=k^2/(2\pi D)$ is the upper edge of the operating window, Eq.~\eqref{eq:window_dim}. For $\avg{Sh}\approx8$ and this $\lP$, the lower edge, $\wstar\approx\avg{Sh}^2$, is about 50 to 400~Hz over the range of $D$ in the table. Vanadium couples were measured on plastic formed carbon~\cite{Yamamura2005}, AQDS on glassy carbon~\cite{Huskinson2014}, the couples of Ref.~\cite{Sun2006} at Pt nanoelectrodes, and Ru(NH$_3$)$_6^{3+/2+}$ of Ref.~\cite{Muzikar2004} on Au.}
    \label{tab:kinetics}
    {\setlength{\tabcolsep}{10pt}%
    \begin{tabular}{@{}lllll@{}}
        \toprule
        Couple & $k^0$ (cm/s) & $D$ ($10^{-9}$ m$^2$/s) & $\Da$ & $f_k$ \\
        \midrule
        V$^{2+}$/V$^{3+}$~\cite{Yamamura2005} & $5.3\times10^{-4}$ & $0.7$ & 0.08 & 6~mHz \\
        VO$^{2+}$/VO$_2^{+}$~\cite{Yamamura2005} & $8.5\times10^{-4}$ & $0.7$ & 0.12 & 16~mHz \\
        AQDS~\cite{Huskinson2014} & $7.2\times10^{-3}$ & 0.38 & 1.9 & 2~Hz \\
        TEMPO$^{b}$~\cite{Stumme2023} & 0.1--0.5 & 2--4 & 2.5--25 & 40~Hz--2~kHz \\
        TCNQ~\cite{Sun2006} & 1.1 & 1.98 & 56 & 10~kHz \\
        Ru(NH$_3$)$_6^{3+/2+}$, Au$^{c}$~\cite{Muzikar2004} & 0.54--1.43 & 0.66--0.51 & 82--280 & 7--64~kHz \\
        FcCH$_2$OH~\cite{Sun2006} & 6.8 & 0.78 & 870 & 0.9~MHz \\
        Ferrocene~\cite{Sun2006} & 8.4 & 2.05 & 410 & 0.5~MHz \\
        Ru(NH$_3$)$_6^{3+/2+}$, Pt~\cite{Sun2006} & 17 & 0.67 & 2500 & 7~MHz \\
        \bottomrule
    \end{tabular}}
    \begin{flushleft}
        \footnotesize $^{a}$Ranges of $\Da$ and $f_k$ are the extreme combinations of the reported ranges of $k^0$ and $D$. $^{b}$In 0.02 to 0.1~M HClO$_4$; each $k^0$ is paired with the $D$ measured in the same electrolyte.
    \end{flushleft}
\end{table*}

\subsection{Application to specific technologies}
\label{sec:applications}

Table~\ref{tab:kinetics} evaluates Eq.~\eqref{eq:window_dim} for representative redox couples and separates the applications into three groups.

For aqueous flow batteries, vanadium redox couples~\cite{Lourenssen2019,Yamamura2005} and the quinone 9,10-anthraquinone-2,6-disulfonic acid (AQDS)~\cite{Huskinson2014} have kinetic frequencies $f_k$ of approximately 1~Hz or lower. These values are far below the lower edge of the operating window, which is approximately 50~Hz. Near equilibrium, these systems are therefore kinetics-limited at high frequency. The spectral Sherwood number saturates at $\Da$, and the Warburg regime is inaccessible regardless of pore size. For these chemistries, the $\Da\to\infty$ results provide an upper bound that can be approached only at substantial overpotential. Increasing $k$ through electrode activation is therefore more beneficial than modifying the microstructure.

Nonaqueous flow batteries provide a more favorable case. TEMPO has fast electron transfer and a positive redox potential~\cite{Suga2004,Wei2014,Zhao2021,Stumme2023}, making it the first group in Table~\ref{tab:kinetics} for which an operating window opens near equilibrium. This window spans up to approximately one decade in frequency. Ferrocene-type couples, which have been paired with TEMPO in nonaqueous catholytes~\cite{Kosswattaarachchi2018,Gong2015}, have kinetic frequencies near 1~MHz at Pt nanoelectrodes~\cite{Sun2006}. Such chemistries could, in principle, use high-frequency transfer to accommodate short power transients.

These estimates require caution. The reported rate constants were obtained using model Pt electrodes in dilute solutions, whereas practical electrodes are often carbon and practical electrolytes are concentrated~\cite{Stumme2023}. The rate constant on the working electrode must therefore be measured under conditions close to those of the intended application before relying on this estimate.

The fastest entries in Table~\ref{tab:kinetics} were measured using methods that rely on the same transport physics considered here. In ac admittance voltammetry, the Warburg contribution must be separated from the charge-transfer resistance~\cite{Muzikar2004,Baranski1991}. A measurement at 94~kHz in 0.1~M HClO$_4$ corresponds to $\omega=5.9\times10^5$~s$^{-1}$, which is close to $k^2/D=4.0\times10^5$~s$^{-1}$. The experiment therefore operates near the upper edge of Eq.~\eqref{eq:window_dim}, where the kinetic and diffusive impedances are comparable and can be separated. Similarly, steady-state voltammograms at nanoelectrodes become essentially reversible when $k^0r_e/D\gtrsim10$~\cite{Sun2006}, where $r_e$ is the electrode radius. This parameter is a Damk\"ohler number. Both observations support the scaling used here.

Reported rate constants for a given redox couple also vary with the electrode and electrolyte. For $\mathrm{Ru(NH_3)_6^{3+/2+}}$, earlier ultramicroelectrode measurements gave values between 0.36 and 0.45~cm/s~\cite{Wipf1988,Birkin1997}, compared with the values listed in Table~\ref{tab:kinetics}. Rate constants should therefore be measured under conditions representative of the intended application. Extending these methods to porous electrodes, as required for flow-battery materials, requires a transfer law that accounts for porous geometry. Equation~\eqref{eq:z_correction} provides such a relation. The geometric correction is purely imaginary, so it leaves the real-axis intercept $1/\Da$ in Eqs.~\eqref{eq:z_Plates} and \eqref{eq:z_Tube} unchanged. Its imaginary part instead measures the curvature and edge contributions of the electrode.

Pulse protocols in electrodeposition and metal recovery already exploit the principle quantified by Eqs.~\eqref{eq:average_bound} and \eqref{eq:Sand}~\cite{Ibl1980,Cheh1971,Mariani2022,ZLiu2023,Tian2025}. The present results extend this principle to porous deposits and porous current collectors. They show that the pulse limit is set by $\avg{Sh}$, and therefore depends on flow and microstructure, whereas the high-frequency flux does not. The deposit geometry changes during operation, however, and this effect is not included in the present model.

\subsection{Electrode-scale limits}

The pore-scale operating window is not the only constraint. At the electrode scale, ionic resistance can confine high-frequency current near the electrode face~\cite{deLevie1963}. Balancing the ionic resistance with the Warburg impedance per unit interfacial area,
$Z_{W,s}=RT/(F^2\avg{c}\sqrt{\omega D})$, gives the penetration depth
\begin{equation}
    L_{\mathrm{pen}}\approx\sqrt{\frac{\sigma Z_{W,s}}{\sA}},
    \label{eq:penetration}
\end{equation}
where $\sigma$ is the electrolyte conductivity. For 50~mM TEMPO in 500~mM TBAPF$_6$, the measured conductivity is approximately 2.5~S/m~\cite{Stumme2023}. With $\sA=7.5\times10^4$~m$^{-1}$ and $f=1$~kHz, Eq.~\eqref{eq:penetration} gives $L_{\mathrm{pen}}\approx7~\mu$m, comparable to a single particle. This is an upper estimate because the effective conductivity of a porous medium is lower than that of the bulk electrolyte. At these frequencies, thick felts would therefore draw current mainly through a thin region near the separator or membrane. Moreover, $L_{\mathrm{pen}}$ decreases as $\omega^{-1/4}$ and $\avg{c}^{-1/2}$. High-frequency transfer is thus best suited to thin porous layers or to the region adjacent to the separator or membrane in a conventional cell.

Double-layer charging imposes a weaker constraint in concentrated electrolytes. For a double-layer capacitance of $C_{\mathrm{dl}}=20~\mu\mathrm{F}/\mathrm{cm}^2$~\cite{Muzikar2004}, the capacitive and Warburg impedances become comparable at
$\omega\approx(F^2\avg{c}\sqrt{D}/C_{\mathrm{dl}}RT)^2$. This frequency is approximately $2\times10^9$~s$^{-1}$ at 50~mM but decreases to about $7\times10^5$~s$^{-1}$ at 1~mM. Double-layer charging is therefore most relevant under dilute analytical conditions.

Other effects can also limit the present results. At sufficiently high frequency, the diffusion length may approach molecular scales. Equation~\eqref{eq:C_general} does not apply locally at particle contacts or fiber junctions, where the pore gap closes, or to surface roughness smaller than $\delta$, which makes the effective interfacial area frequency-dependent. Straight fibers most closely resemble the rods considered here, making carbon felts a natural first experimental system, provided that measurements are performed away from fiber junctions.
    \section{Conclusions}

We examined frequency-dependent interfacial mass transfer in parallel plates, circular tubes, and twelve porous microstructures that vary in particle shape, symmetry, porosity, flow strength and orientation, and array arrangement. In the diffusion-limited limit, $\Da\to\infty$, every geometry approaches $\Shnd\approx\sqrt{\iu\wstar}$, with magnitude proportional to $\omega^{1/2}$, phase $\pi/4$, leading prefactor 1, and the corresponding semi-infinite Warburg impedance. A boundary-layer analysis shows that this behavior results from the local balance between transient accumulation and diffusion normal to the interface.

We also derived the first correction to this limit for an arbitrary solid. This correction is a constant determined by the interfacial area per pore volume, mean curvature, and sharp edges. It predicts the offset for eleven two-dimensional microstructures without fitted parameters and agrees with an exact solution for spheres. It explains the slow approach of low-porosity media to the Warburg limit, while its storage-dependent contribution cancels from the first correction to the impedance. Published impedance data for a carbon-felt fiber, converted to the present variables, follow the predicted $\Shnd$ and give an offset of $0.94\pm0.13$, consistent with the curvature contribution. Data for a vanadium flow-battery electrode show a pore-scale transport phase of $30$--$38^\circ$, below the Warburg limit, and indicate that the electrode is primarily kinetics-limited over this frequency range.

Because the transport problem is linear, the time-averaged flux remains bounded by the steady limit, while the enhancement applies to the peak flux. The transition-time and operating-window results determine the pulse durations, frequencies, and kinetic rates required to access this enhancement. The upper frequency limit, $\omega\ll k^2/D$, is independent of pore size. Applied to reported rate constants, this criterion shows that vanadium and quinone couples near equilibrium lie outside the operating window, TEMPO catholytes lie near its lower edge, and ferrocene-type couples lie well within it on model electrodes. At the electrode scale, ionic resistance further confines the response to a thin region near the electrode face.

These results guide the selection of redox chemistries and electrode thicknesses for short power transients in flow batteries, establish pulse limits for electrodeposition and metal recovery in porous media, and extend Warburg-based kinetic measurements to porous electrodes. They also motivate three-dimensional simulations, finite-$\Da$ calculations for porous geometries, and experiments that measure rate constants and the storage contribution to $C$ on working carbon electrodes under high-frequency and pulsed operation.

\section*{Acknowledgments}

The U.S. Office of Naval Research (Award No.~N00014-22-1-2577) and the U.S. National Science Foundation (Award No.~1931659) supported this research. Graduate student support from the Department of Mechanical Science and Engineering at the University of Illinois Urbana--Champaign is also acknowledged.

\section*{Data availability}

The data supporting the findings of this article are not publicly available but can be obtained from the authors upon reasonable request.

    \bibliographystyle{apsrev4-2} % Adds explicit rule for revtex4-2
    \bibliography{Z01_References}% Produces the bibliography via BibTeX.

    \appendix %%%%%%%%%%%%%%%%%%%%%%%%%%%%%%%%%%%%%%%%%%%%%%%%%%%%%%%%%%%%%%%%%%%%%%%

\section{\label{ap:A}Impedance from the Spectral Sherwood Number}

This Appendix derives the relation between the spectral Sherwood number and the electrochemical impedance. We consider one species of a redox couple with negligible electromigration. The second species is treated analogously and contributes an additional impedance in series. We assume a one-electron reaction and equal molecular diffusivities and spectral Sherwood numbers for the two species.

Let $\tc_{s,eq}$ denote the interfacial concentration required by the Nernst equation, and let $c^0_{s,eq}$ denote its initial equilibrium value. A perturbation of the interfacial concentration produces a corresponding perturbation of the electrode potential. Linearizing the Nernst equation about the initial equilibrium state gives
\begin{equation}
    \Delta\ft{E}=\mp\frac{RT}{F}\left(\frac{\tc_{s,eq}}{c^0_{s,eq}}\right),
    \label{eq:A1}
\end{equation}
where the upper sign applies to the reduced species and the lower sign to the oxidized species. Thus, the voltage perturbation is proportional to the relative perturbation of the equilibrium interfacial concentration. The total voltage perturbation is the sum of the contributions from the two species, so their impedance contributions add in series.

The interfacial flux $\ft{\jj}_s$ is defined as positive from the interface into the fluid. With this convention, Faraday's law relates the species flux to the electric current, while the spectral Sherwood number relates the same flux to the concentration difference between the equilibrium interfacial concentration and the intrinsic fluid-phase average:
\begin{equation}
    \ft{I}=-\frac{\Atot F}{\nu}\avg{\ft{\jj}_s},
    \qquad
    \avg{\ft{\jj}_s}=\frac{D}{\lP}\Shnd\left(\tc_{s,eq}-\avg{\tc}\right),
    \label{eq:A2}
\end{equation}
where $\nu$ is the signed stoichiometric coefficient of the species. The first relation converts the molar flux into electric current, and the second expresses the flux through the spectral mass-transfer coefficient.

The area-specific impedance of this species is defined as ${\Zeff}=\Atot\,\Delta\ft{E}/\ft{I}$. Substituting Eqs.~\eqref{eq:A1} and \eqref{eq:A2} gives
\begin{equation}
    {\Zeff}=\left(\frac{|\nu|\lP RT}{F^2Dc^0_{s,eq}}\right)\,\frac{1}{\Shnd(1-\Cnd)},
    \qquad
    \Cnd\equiv\frac{\avg{\tc}}{\tc_{s,eq}},
    \label{eq:A3}
\end{equation}
where the sign in Eq.~\eqref{eq:A1} cancels the sign of $\nu$. The ratio $\Cnd$ measures how closely the pore-fluid concentration follows the equilibrium interfacial concentration. Thus, $1-\Cnd$ is the normalized concentration difference that drives interfacial mass transfer.

For a symmetric redox couple with equal initial concentrations, the oxidized and reduced species contribute equally. The total impedance is obtained by adding their contributions. Here, we normalize the impedance contribution of one species by the prefactor in Eq.~\eqref{eq:A3}, giving
\begin{equation}
    \Znd=\frac{1}{\Shnd(1-\Cnd)}.
    \label{eq:A4}
\end{equation}
Equation~\eqref{eq:A4} contains two factors. The spectral Sherwood number describes interfacial mass-transfer resistance, while $1-\Cnd$ describes the concentration difference required to sustain the transfer.

We determine $\Cnd$ from the volume-averaged mass balance. The term $\iu\omega\Vf\avg{\tc}$ represents oscillatory solute accumulation in the fluid volume. This accumulation equals the total interfacial flux entering the fluid, $\Atot\avg{\ft{\jj}_s}$. As in the main text, we assume that the macroscopic concentration is locally uniform, so neither advection nor macroscopic diffusion contributes to this balance:
\begin{equation}
    \iu\omega\Vf\avg{\tc}=\Atot\avg{\ft{\jj}_s}.
\label{eq:A5}
\end{equation}
Using the flux relation in Eq.~\eqref{eq:A2}, together with the definitions of $\Shnd$ and $\sA=\Atot/\VRVE$, gives
\begin{equation}
    \Cnd=\left[1+\frac{\eps\,\iu\wstar}{\sA\lP\Shnd}\right]^{-1}.
    \label{eq:A6}
\end{equation}
The term involving $\iu\wstar$ represents temporal solute storage in the fluid volume, while the term involving $\sA\lP\Shnd$ represents interfacial mass transfer.

Substituting Eq.~\eqref{eq:A6} into Eq.~\eqref{eq:A4} separates the impedance into mass-transfer and storage contributions:
\[
    \Znd=\frac{1}{\Shnd}+\frac{\sA\lP}{\eps\,\iu\wstar}.
\]
This is Eq.~\eqref{eq:ZSh}. For parallel plates and circular tubes, the corresponding one-dimensional closure solutions give Eqs.~\eqref{eq:z_Plates} and \eqref{eq:z_Tube}.

\section{\label{ap:B}First Correction to the High-Frequency Limit}

This Appendix derives Eq.~\eqref{eq:C_general}. In the perfectly reactive limit, $\Da\to\infty$, we write
$a=a_\infty+(1-a_\infty)\Phi$, where $a_\infty=\alpha/(\iu\omega)$ and $\Phi=1$ on $s$. Under the conditions in Eq.~\eqref{eq:C_conditions}, advection is negligible at the order considered here. Thus, $\Phi$ satisfies
$\iu\omega\Phi=D\nabla^2\Phi$ and decays away from the interface.

Near a smooth surface, let $n$ denote the distance from the interface into the fluid and define
$m=\sqrt{\iu\omega/D}$. The Laplacian in the normal direction is
$\nabla^2\Phi\approx\partial_n^2\Phi+2\rho\partial_n\Phi$, where $\rho>0$ for a surface convex toward the fluid. Setting $\Phi=e^{-mn}\chi(n)$ with $\chi(0)=1$ and retaining the two leading terms gives $\chi'=-\rho\chi$. Therefore,
\begin{equation}
    -\partial_n\Phi\big|_{n=0}=m+\rho+O(m^{-1}).
    \label{eq:B_grad}
\end{equation}

The zero-mean constraint $\avg{a}=0$ determines the far-field value:
\begin{equation}
    a_\infty=-\frac{\avg{\Phi}}{1-\avg{\Phi}},\qquad\avg{\Phi}\approx\frac{\sA}{\eps m}.
    \label{eq:B_ainf}
\end{equation}
The approximation for $\avg{\Phi}$ follows by integrating $e^{-mn}$ across the boundary layer over the interfacial area $\Atot$ and dividing by the fluid volume $\Vf$. Substituting Eqs.~\eqref{eq:B_grad} and \eqref{eq:B_ainf} into Eq.~\eqref{eq:Sh} gives
\begin{equation}
    \Shnd=(1-a_\infty)\lP\left(m+\avg{\rho}_s\right)\approx\lP m+\frac{\sA\lP}{\eps}+\lP\avg{\rho}_s.
    \label{eq:B_result}
\end{equation}
Here, we used $\eps/(\sA\Vf)=1/\Atot$. The factor $m$ cancels between $a_\infty$ and the leading wall-normal gradient, making the storage contribution real. The curvature contribution is also real, which explains why the constant $C$ in Eq.~\eqref{eq:C_general} is real. The neglected terms are of order $1/(\lP m)=1/\sqrt{\iu\wstar}$ and are therefore complex, as stated in Eq.~\eqref{eq:offset_expansion}. This derivation assumes that the diffusion length $\delta=1/|m|$ is small compared with the local radius of curvature, the narrowest pore dimension, and $\lP$.

Near a sharp edge, the local problem is described by a two-dimensional wedge solution. The short-time heat content of a polygon with Dirichlet edges is
$|\mathcal{P}|-(2/\sqrt{\pi})|\partial\mathcal{P}|\,t^{1/2}+\sum_e\Lambda(\gamma_e)\,t$~\cite{vandenBerg1990}, where $|\mathcal{P}|$ and $|\partial\mathcal{P}|$ are the area and perimeter of the polygon, $t$ is time, and $\Lambda$ is defined in Eq.~\eqref{eq:Lambda_gamma}. Taking the Laplace transform gives a transfer function proportional to
$|\partial\mathcal{P}|m-\sum_e\Lambda(\gamma_e)$. Thus, each edge contributes
$-\Lambda(\gamma_e)L_e\lP/\Atot$ to $\Shnd$, producing the final term in Eq.~\eqref{eq:C_general}. Vertices in three dimensions enter only at higher order. A corner rounded at a radius much larger than $\delta$ contributes through $\avg{\rho}_s$ instead and therefore gives a different coefficient. For a right-angled convex corner, the coefficient of $\lP/\Atot$ is $\pi/4$ when the corner is rounded and $0.602$ when it is sharp.

The result reproduces the exact solutions for the canonical geometries. Expanding Eqs.~\eqref{eq:z_Plates} and \eqref{eq:z_Tube} gives $C=1$ for parallel plates and $C=3/2$ for circular tubes. These values agree with Eq.~\eqref{eq:C_general}. For parallel plates, $\sA\lP/\eps=1$ and $\rho=0$. For circular tubes, $\sA\lP/\eps=2$ and $\lP\rho=-1/2$, because the fluid lies on the concave side of the interface. We confirmed the corresponding two-dimensional values numerically to within 0.02 at $\wstar=10^4$.

The two reference solutions used in Sec.~\ref{sec:experiments} follow from the same closure problem. Consider first a single cylinder of radius $r_0=d/2$ in an unbounded fluid without flow. Because the fluid volume is unbounded, $\avg{\Phi}\to0$ and $a_\infty\to0$. The solution of $\iu\omega a=D\nabla^2a$ that equals 1 at $r=r_0$ and decays as $r\to\infty$ is
$a=K_0(mr)/K_0(mr_0)$. Inserting its wall-normal gradient into Eq.~\eqref{eq:Sh} with $\lP=d$ gives Eq.~\eqref{eq:Sh_fiber}. Since
$K_1(y)/K_0(y)=1+1/(2y)+O(y^{-2})$ for large $y$, Eq.~\eqref{eq:Sh_fiber} approaches $x+1$. This agrees with Eq.~\eqref{eq:C_general}, for which $\sA\lP/\eps=0$ and $\lP\avg{\rho}_s=1$.

For a planar layer of thickness $\lP$ with $a=1$ at the reactive surface and $a=0$ at the opposite boundary,
$a=\sinh[m(\lP-n)]/\sinh(m\lP)$. Its wall-normal gradient gives Eq.~\eqref{eq:Sh_planar}. This solution approaches $x$ without a constant offset because the far boundary has a fixed concentration, so there is no storage contribution, and the reactive surface is flat.

In three dimensions, we solved the closure problem exactly for a sphere of diameter $d$ centered within a concentric spherical fluid cell with a no-flux outer boundary, using $\lP=d$. Table~\ref{tab:3} compares the exact offset with the prediction
$C=6(1-\eps)/\eps+2$. The difference decreases with increasing frequency, as expected for a next-order correction.

\begin{table}[t!]
    \centering
    \caption{Exact offset for a sphere of diameter $d=\lP$ at the center of a spherical cell with a no-flux outer boundary, for $\Da\to\infty$ and $\Pe=0$, compared with $C=6(1-\eps)/\eps+2$ from Eq.~\eqref{eq:C_general}. The difference between the real part and $C$ decreases by a factor of $\sqrt{10}$ for each decade in $\wstar$, as expected for the next term, of order $(\wstar)^{-1/2}$, in Eq.~\eqref{eq:offset_expansion}, and the imaginary part vanishes at the same rate.}
    \label{tab:3}
    \begin{tabular*}{\columnwidth}{@{\extracolsep{\fill}}ccccc}
        \toprule
        & & \multicolumn{2}{c}{$\Shnd-\sqrt{\iu\wstar}$} & \\
        \cmidrule(lr){3-4}
        $\eps$ & $\wstar$ & Re & Im & $C$ \\
        \midrule
        0.875 & $4\times10^3$ & 2.904 & $-0.049$ & 2.857 \\
        0.875 & $4\times10^4$ & 2.872 & $-0.015$ & 2.857 \\
        0.875 & $4\times10^5$ & 2.862 & $-0.005$ & 2.857 \\
        \midrule
        0.545 & $4\times10^3$ & 7.507 & $-0.584$ & 7.009 \\
        0.545 & $4\times10^4$ & 7.169 & $-0.167$ & 7.009 \\
        0.545 & $4\times10^5$ & 7.060 & $-0.051$ & 7.009 \\
        \bottomrule
    \end{tabular*}
\end{table}

\section{\label{ap:C}Nomenclature}

Symbols are listed in the order Roman, Greek, and other notation. Subscripts and superscripts that are not listed separately are defined with the symbol to which they are attached.

\newenvironment{nomlist}{%
    \begin{list}{}{%
        \setlength{\labelwidth}{1.9cm}%
        \setlength{\labelsep}{0.2cm}%
        \setlength{\leftmargin}{2.1cm}%
        \setlength{\itemsep}{1pt}%
        \setlength{\parsep}{0pt}%
        \setlength{\topsep}{2pt}%
        \renewcommand{\makelabel}[1]{##1\hfil}}%
    \small\rightskip=0pt plus 2em\relax}{\end{list}}

\subsection*{Roman symbols}
\begin{nomlist}
    \item[$a$] Closure variable for the interfacial concentration difference
    \item[$a_\infty$] Uniform far-field value of $a$
    \item[$\Atot$] Total solid--liquid interfacial area in the RVE (perimeter per unit depth in 2D)
    \item[$\bcv$] Closure variable for the macroscopic concentration gradient
    \item[$C$] First correction to the Warburg limit, Eq.~\eqref{eq:C_general}
    \item[$C_1$] Coefficient of the next-order term, Eq.~\eqref{eq:offset_expansion}
    \item[$C_{\mathrm{dl}}$] Double-layer capacitance
    \item[$c$] Concentration
    \item[$\tc$] Fourier-transformed concentration
    \item[$\avg{c}$] Intrinsic fluid-phase average concentration
    \item[$c_{s,eq}$] Interfacial concentration in equilibrium with the electrode potential
    \item[$c^0_{s,eq}$] Initial value of $c_{s,eq}$
    \item[$\Delta c_{\mathrm{av}}$] Time average of $c_{s,eq}-\avg{c}$
    \item[$D$] Molecular diffusivity
    \item[$D_{BL}$] Effective diffusivity in the boundary-layer model of Ref.~\cite{Vivona2020}
    \item[$d$] Fiber or sphere diameter
    \item[$d_P$] Equivalent particle diameter, $4V_s/\Atot$
    \item[$d_{BL}$] Boundary-layer thickness in the model of Ref.~\cite{Vivona2020}
    \item[$\Da$] Damk\"ohler number, $k\lP/D$
    \item[$\Da_c$] Damk\"ohler number of the redox couple, Eq.~\eqref{eq:LL_conversion}
    \item[$\Da_{\mathrm{eff}}$] Effective Damk\"ohler number, Eq.~\eqref{eq:V_conversion}
    \item[$\Delta\ft{E}$] Perturbation of the electrode potential
    \item[$F$] Faraday constant
    \item[$f$] Frequency
    \item[$f_k$] Kinetic frequency, $k^2/(2\pi D)$
    \item[$g_{\min}$] Narrowest local pore dimension
    \item[$h$] Half-spacing of parallel plates
    \item[$I_0$, $I_1$] Modified Bessel functions of the first kind
    \item[$\ft{I}$] Fourier-transformed electric current
    \item[$J$] Pulse flux as a multiple of $j_{\mathrm{lim}}$
    \item[$K_0$, $K_1$] Modified Bessel functions of the second kind
    \item[$k$] Interfacial reaction-rate constant
    \item[$k^0$] Standard rate constant
    \item[$k_f$, $k_b$] Forward and backward rate constants
    \item[$L_e$] Length of sharp edge $e$
    \item[$L_{\mathrm{pen}}$] Ionic penetration depth, Eq.~\eqref{eq:penetration}
    \item[$\lP$] Characteristic pore length
    \item[$\lRVE$] Side length of the RVE
    \item[$l_{\text{ref}}$] Reference length in Table~\ref{tab:1}
    \item[$l_w$] Mean width of a convex solid
    \item[$M$] Number of separate inclusions per unit cell
    \item[$m$] Complex wavenumber, $\sqrt{\iu\omega/D}$
    \item[$\mathbf{j}_s$] Interfacial molar flux, positive from the interface into the fluid
    \item[$j_{\mathrm{av}}$] Time-averaged interfacial flux
    \item[$j_{\mathrm{lim}}$] Steady limiting flux
    \item[$j_{\mathrm{peak}}$] Peak interfacial flux
    \item[$n$] Distance from the interface into the fluid
    \item[$\nhat$] Unit normal, pointing from the fluid into the solid
    \item[$p$, $q$] Geometric dimensions of the inclusions, Fig.~\ref{fig:2}
    \item[$\Pe$] P\'eclet number, $|\mathbf{u}_s|d_P/D$
    \item[$R$] Gas constant
    \item[$R_c$] Radius of curvature of a surface feature
    \item[$R_u$] Series (uncompensated) resistance
    \item[$R_{ct}$] Charge-transfer resistance
    \item[$R_\Omega$] Ohmic resistance
    \item[$R_{HF}$] High-frequency intercept, $R_\Omega+R_{ct}$
    \item[$r$] Radial coordinate
    \item[$r_0$] Fiber or sphere radius
    \item[$r_t$] Tube radius
    \item[$r_e$] Nanoelectrode radius
    \item[$s$] Solid--liquid interface
    \item[$\sA$] Interfacial area per unit RVE volume, $\Atot/\VRVE$
    \item[$\avg{Sh}$] Pseudo-steady Sherwood number
    \item[$\Shnd$] Spectral Sherwood number, Eq.~\eqref{eq:Sh}
    \item[$T$] Temperature
    \item[$t$] Time
    \item[$\uu$] Velocity field
    \item[$u_\tau$, $u_n$] Tangential and normal velocity components
    \item[$\mathbf{u}_s$] Superficial (Darcy) velocity, $\eps\avg{\uu}$
    \item[$\VRVE$] Volume of the RVE
    \item[$\Vf$, $\Vs$] Fluid and solid volumes in the RVE (areas per unit depth in 2D)
    \item[$W(\xi)$] Bessel-function ratio in Eq.~\eqref{eq:LL_ZF}
    \item[$x$] Complex frequency variable, $\sqrt{\iu\wstar}$
    \item[$\xl$] Pore-scale position vector
    \item[$Y_0$] Constant-phase-element coefficient
    \item[$Z$] Measured impedance
    \item[$Z_F$] Faradaic impedance
    \item[$Z_W$] Mass-transfer impedance in the data of Ref.~\cite{Vivona2020}
    \item[$Z_{W,s}$] Warburg impedance per unit interfacial area
    \item[$\Zeff$] Area-specific diffusion impedance of one species
    \item[$\Znd$] Nondimensional diffusion impedance, Eq.~\eqref{eq:ZSh}
\end{nomlist}

\subsection*{Greek symbols}
\begin{nomlist}
    \item[$\alpha$] Uniform source term in the closure problem
    \item[$\beta$] Interior angle of the solid at a sharp edge
    \item[$\gamma$, $\gamma_e$] Fluid-side angle of a sharp edge, $2\pi-\beta$
    \item[$\delta$] Diffusion length, $\sqrt{D/\omega}=\lP/\sqrt{\wstar}$
    \item[$\eps$] Porosity, $\Vf/\VRVE$
    \item[$\eta$] Stretched normal coordinate, $n/\delta$
    \item[$\theta$] Flow angle
    \item[$\theta_{\Shnd}$] Phase of $\Shnd$
    \item[$\kappa$] Curvature of a two-dimensional boundary
    \item[$\kappa_1$, $\kappa_2$] Principal curvatures of a surface    
    \item[$\Lambda(\gamma)$] Edge coefficient, Eq.~\eqref{eq:Lambda_gamma}
    \item[$\nu$] Signed stoichiometric coefficient
    \item[$\xi$] Bessel argument in Eq.~\eqref{eq:LL_ZF}, $\sqrt{\iu\omega\tau_d}$
    \item[$\Cnd$] Ratio of average to equilibrium interfacial concentration, $\avg{\tc}/\tc_{s,eq}$
    \item[$\rho$] Mean curvature, $(\kappa_1+\kappa_2)/2$, positive where the solid is convex
    \item[$\avg{\rho}_s$] Area average of $\rho$ over the smooth part of the interface
    \item[$\sigma$] Electrolyte conductivity
    \item[$\tau$] Sand transition time
    \item[$\tau^*$] Dimensionless transition time, $\tau D/\lP^2$
    \item[$\tau_d$] Diffusion time of the fiber, $r_0^2/D$
    \item[$\Phi$] Boundary-layer part of $a$
    \item[$\phi$] Constant-phase-element exponent
    \item[$\chi$] Curvature correction to $\Phi$, Appendix~\ref{ap:B}
    \item[$\Psi$] Stream function
    \item[$\psi$] Kinetic parameter of Ref.~\cite{LandonLane2019}, $(k_f+k_b)/\sqrt{D}$
    \item[$\omega$] Angular frequency
    \item[$\wstar$] Dimensionless frequency, $\omega\lP^2/D$
    \item[$\wstar_c$] Crossover frequency, of order $\avg{Sh}^2$
\end{nomlist}

\subsection*{Other notation}
\begin{nomlist}
    \item[$\iu$] Imaginary unit, $\iu^2=-1$
    \item[$\ft{(\cdot)}$] Fourier-transformed quantity
    \item[$\tf{(\cdot)}$] Frequency-dependent averaged quantity
    \item[$\avg{\cdot}$] Intrinsic fluid-phase average
    \item[$\avg{\cdot}_s$] Average over the interfacial area
    \item[$\mathcal{P}$] Polygon in the heat-content expansion, Appendix~\ref{ap:B}
    \item[$\mathcal{R}$] Higher-order residual in the measured offset
    \item[$e$] Index of a sharp edge
    \item[$\zeta$, $y$] Dummy variables in Eq.~\eqref{eq:Lambda_gamma} and Appendix~\ref{ap:B}
\end{nomlist}

\subsection*{Abbreviations}
\begin{nomlist}
    \item[AC] Alternating current
    \item[AQDS] 9,10-Anthraquinone-2,6-disulfonic acid
    \item[BL] Boundary layer
    \item[CPE] Constant-phase element
    \item[HF] High frequency
    \item[OLC] Over-limiting current
    \item[RFB] Redox-flow battery
    \item[RVE] Representative volume element
    \item[TEMPO] (2,2,6,6-Tetramethylpiperidin-1-yl)oxyl
    \item[TBAPF$_6$] Tetrabutylammonium hexafluorophosphate
\end{nomlist}
\end{document}